\documentclass[12pt]{article}

\usepackage[usenames]{color}
\usepackage{setspace}

\usepackage{graphicx}
\usepackage{amsmath}
\usepackage{amssymb}
\usepackage{amsthm}

\usepackage{cite}

\newcommand{\fun}[1]{\!\left({#1}\right)}
\newcommand{\funal}[1]{\!\left[#1\right]}

\newcommand{\VEV}[1]{\left\langle #1 \right\rangle}
\newcommand{\del}{\partial}

\newcommand{\nn}{\nonumber}
\newcommand{\diag}{{\rm diag.}}

\newcommand{\order}[1]{\mathcal{O}(#1)} 
\newcommand{\M}[1]{M_{\mathrm{#1}}}
\newcommand{\Eq}[1]{Eq.~(\ref{#1})} 

\newcommand{\red}[1]{{\color{Red}{#1}}}

 \renewcommand{\red}[1]{{#1}}

\newcommand{\Zex}{{\mathbb Z}_2^{ex}}
\newcommand{\Zp}{{\mathbb Z}_2^{5d}}
\newcommand{\Zcomb}{{\mathbb Z}_2^{comb}}

\newcommand{\modd}{m_-}
\renewcommand{\modd}{m}

\newcommand{\mKK}{M}

\newcommand{\Z}[1]{{\mathbb Z}_#1}
\newcommand{\abs}[1]{\left| #1 \right|}

\newcommand{\supP}[1]{^{(#1)}}
\newcommand{\sla}[1]{#1\!\!\!\!/}

\newcommand{\bequ}{\begin{equation}}
\newcommand{\eequ}{\end{equation}}
\newcommand{\beqn}{\begin{eqnarray}}
\newcommand{\eeqn}{\end{eqnarray}}

\newcommand{\Ls}{\left(}
\newcommand{\Rs}{\right)}
\newcommand{\Lm}{\left\{}
\newcommand{\Rm}{\right\}}

\newcommand{\RR}{\right.}

\newcommand{\lsim}{%
\raise0.3ex\hbox{$\;<$\kern-0.75em\raise-1.1ex\hbox{$\sim\;$}}}
\newcommand{\gsim}{%
\raise0.3ex\hbox{$\;>$\kern-0.75em\raise-1.1ex\hbox{$\sim\;$}}}

\def\g5D{g_{5D}}
\def\gD{g_{D}}

\def\Hc{\mathrm{H.c.}}

\begin{document}

\begin{flushright}
\end{flushright}

\begin{center}
{\LARGE\bf 
Dirac gaugino 
from\\ grand gauge-Higgs unification}

\vskip 1.4cm

{\large  
Hiroaki~Nakano$^{a,}$\footnote{E-mail: nakano@muse.sc.niigata-u.ac.jp}, 
Masamichi~Sato$^{b,}$\footnote{E-mail: masamichi.sato@muse.sc.niigata-u.ac.jp}, 
Osamu~Seto$^{c,d,}$\footnote{E-mail: seto@particle.sci.hokudai.ac.jp}, 
\\[5pt]
and Toshifumi~Yamashita$^{e,}$\footnote{E-mail: tyamashi@aichi-med-u.ac.jp}
}
\\
\vskip 1.0cm
{\it $^a$ Department of Physics, Niigata University, Niigata, 950-2181, Japan\\
$^b$ Graduate School of Science and Technology, Niigata University, Niigata,\\ 
950-2181, Japan\\
$^c$ Institute for the Advancement of Higher Education, Hokkaido University,\\
 Sapporo 060-0817, Japan\\ 
$^d$Department of Physics, Hokkaido University, Sapporo 060-0810, Japan\\
$^e$ Department of Physics, Aichi Medical University, Nagakute 480-1195, Japan\\
}

\vskip 1.5cm

\begin{abstract}
We show that 
models of the Dirac gaugino can naturally be embedded
into a kind of the grand unified theory (GUT), 
the grand gauge-Higgs unification (gGHU) model,
with the gauge group $SU(5)\times SU(5)/\mathbb{Z}_2$ 
on an $S^1/\mathbb{Z}_2$ orbifold.
The supersymmetric gGHU is known to posess 
a light chiral adjoint supermultiplet 
after the GUT breaking, 
thank to the exchange symmetry of two $SU(5)$ groups.
Identifying the `predicted' adjoint fermion
with the Dirac partner of the gaugino,
we argue that
the supersoft term, responsible for the Dirac gaugino mass,
can be obtained from 
the supersymmetric Chern-Simons (CS) like term 
in the gGHU setup.
Although the latter term does not respect the exchange symmetry,
we propose a novel way to introduce its breaking effect 
within a consistent orbifold construction.
We also give a concrete setup of fermion field contents
(bulk and boundary-localized fermions)
that induce the requisite CS-like term,
and calculate its coefficient 
from the bulk profile of chiral fermion zero modes.
Our gGHU setup may be regarded as an extra-dimensional realization
of the Goldstone gaugino scenario that was proposed before
as a solution to the problem of the adjoint scalar masses.
\end{abstract}
\end{center}

\vskip 1.0 cm

\newpage



%
\section{Introduction}
\label{Sec:introduction}
%

The standard model (SM) of the particle physics, 
with a possibly simple extension for the neutrino masses,
is an extremely good phenomenological model. 
It basically explains the vast amounts of the experimental results 
below the TeV scale. 
Given the excellent phenomenological success, 
it may be suggestive to extrapolate the model to 
the very high energy region never reached by the experiments. 
Such a na\"{i}ve extrapolation indicates~\cite{%
Degrassi:2012ry,Buttazzo:2013uya,Bezrukov:2012sa}
that the quartic coupling of the Higgs field vanishes 
at an intermediate scale around $10^{11}\,\mathrm{GeV}$. 
It is interesting to assume that 
this is a footprint of the new physics beyond the SM. 
So far,
two scenarios have been proposed as such candidates
that predict the vanishing of the quartic coupling:
the gauge-Higgs unification scenario~\cite{GHU1,GHU2,GHU3,GHU-lambda} 
and the Dirac gaugino scenario~\cite{%
DiracGaugino,Polchinski:1982an,Unwin:2012fj,Fox:2014moa}.

In the former extra-dimensional scenario~\cite{GHU1,GHU2,GHU3}, 
the electroweak (EW) gauge symmetry is broken 
via the so-called Hosotani mechanism~\cite{%
Hosotani1,Hosotani2,Hosotani3,Hosotani4}, 
in which a gauge field in higher dimensions
gives rise to the zero mode in its extra-dimensional components
that takes ``nontrivial'' vacuum expectation values (VEVs).
In other words, the Higgs field is a part of the gauge field 
and thus has the vanishing self-coupling above the scale 
where the extra dimensions become visible. 
This can be expressed as a boundary condition, 
named the ``gauge-Higgs condition"~\cite{GHC1,GHC2}, 
on the renormalization group equation 
of the Higgs quartic coupling 
in the four-dimensional (4D) effective theory. 
It requires the coupling constant vanishing 
at the compactification scale, which is
to be identified with the intermediate scale~\cite{GHU-lambda}.

In the latter supersymmetric (SUSY) scenario~\cite{%
DiracGaugino,Polchinski:1982an}, 
adjoint chiral supermultiplets are introduced so that 
the gauginos are (pseudo-)Dirac fermions instead of Majorana. 
In the pure-Dirac limit, the $D$-term contribution 
to the quartic scalar couplings are canceled 
by the exchange of the scalar component of the adjoint multiplets. 
Then the above intermediate scale may be identified with 
the adjoint scalar mass scale~\cite{Unwin:2012fj, Fox:2014moa}.
%
Aside from this intermediate scale scenario,
the Dirac gaugino models have been studied 
also in the context of the TeV-scale SUSY,
which features other attractive properties of the Dirac gaugino models,
such as the ``supersoftness''~\cite{supersoft} 
and the ``supersafeness''~\cite{supersafe1, supersafe2}.
Given the null results for the signal beyond the SM at the LHC,
the supersafeness property may be helpful 
for relaxing the constraints on 
the SUSY breaking scale~\cite{DG-LHC,Goodsell:2020lpx}. 
The origin of the supersoft operator,
responsible for the Dirac mass term of the gauginos,
and related problems were discussed
in Refs.~\cite{Carpenter:2010as, Csaki:2013fla, GoldstoneGaugino1, GoldstoneGaugino2}.
The issue of the $D$-term cancellation and the Higgs mass
was also addressed, for instance 
in the minimal $R$-symmetric model~\cite{%
Bertuzzo:2014bwa,Diessner:2014ksa,Diessner:2015yna}
and also in the next-to-minimal extension~\cite{Nakano:2015gws}.

The Dirac gaugino scenario is attractive 
as a low-energy effective theory,
but 
it contains some nontrivial assumptions 
to be addressed if we try to construct a concrete UV completion.
%
See \red{Sect.}~\ref{Sec:DiracGaugino} 
for a brief review in this point.
%
Among others,
the required adjoint chiral superfields look less natural
especially when we try to embed the Dirac gaugino models
into a grand unified theory (GUT)~%
\cite{SU(5)GG,GUT1,GUT2,GUT3,GUT4}.

In this respect, 
there is an interesting class of GUT models 
that naturally ``predicts'' the presence of 
light adjoint chiral multiplet:
it is (a version of)
 the grand gauge-Higgs unification 
(gGHU) model~\cite{gGHU1,gGHU2,gGHU-DTS,gGHU-pheno}.\footnote{
Other versions of ``gGHU'' were proposed 
in several contexts
in Refs.~\cite{Lim:2007jv,Hosotani:2015hoa,%
Furui:2016owe,Maru:2019lit,Maru:2019bjr}
and also in Ref.~\cite{Kojima:2017qbt}, where 
the $SU(5)$ symmetry is broken 
by orbifold boundary conditions~\cite{%
orbifoldGUTs1,orbifoldGUTs2,orbifoldGUTs3,%
orbifoldGUTs4,orbifoldGUTs5,orbifoldGUTs6%
},
while one utilizes the Hosotani mechanism to break 
the EW symmetry~\cite{Lim:2007jv,Hosotani:2015hoa,%
Furui:2016owe,Maru:2019lit,Maru:2019bjr}
or to reduce the rank of unified gauge groups~\cite{Kojima:2017qbt}.
}
In the gGHU scenario, 
we utilize the Hosotani mechanism
to break the Georgi-Glashow's $SU(5)_G$ gauge symmetry~\cite{SU(5)GG},
instead of the EW gauge symmetry. 
In this case,
the adjoint Higgs field is identified with the zero mode of 
the extra-dimensional component of the gauge field.
Since such a component has a flat potential at tree level,
the position of the vacuum is determined by quantum corrections.
In general 
the mass and potential of the zero modes would be much distorted
by large radiative corrections
of order of the compactification scale.
In a supersymmetric version of the model, however,
the mass of the zero mode will be of order of 
SUSY breaking scale $\M{SB}$,
which can be much smaller than the compactification scale.
Therefore the existence of the light adjoint chiral superfields 
is a generic prediction~\cite{gGHU-DTS} of 
the gGHU models with supersymmetry\rlap.\footnote{
Phenomenological implications of the chiral adjoints at TeV scale 
were studied in Refs.~\cite{gGHU-DTS,gGHU-pheno},
where characteristic signatures to be observed 
in future collider experiments were also discussed.
} 
In this way, 
such models provide a natural starting point
for constructing a satisfactory UV completion of the Dirac gaugino models.

The purpose of the present article is to show that
the gGHU setup can give 
a good UV completion of the Dirac gaugino models.
Specifically we will show that
the operator responsible for the Dirac gaugino mass 
can be generated as a kind of 
supersymmetric Chern-Simons (CS) term~\cite{HigherDimSUSY}.
Actually we will focus on its bosonic components 
and elaborate how its coefficient can be computed
from a suitable choice of bulk and boundary-localized fermions
and their mass parameters.


In principle one could add 
the requisite CS-like term
to the starting five-dimensional (5D) theory  
by hand.
A more interesting possibility is to start with a 5D theory
without such term and to generate it radiatively.
Actually in the present paper, 
we will be interested in the situation in which
the requisite CS term is generated 
as the term representing anomaly inflow~\cite{Callan:1984sa},
and thus its coefficient can be determined 
through a profile of fermion zero modes spread 
in the 5D bulk.
Alternatively we can calculate it 
by summing up massive Kaluza-Klein (KK) modes.
Such calculation will be applicable
even when no fermion zero mode is present,
as we shall show in a separate publication.





Before going into detailed discussion,
let us summarize here our gGHU setup for Dirac gaugino.
For concreteness, 
we consider a 5D supersymmetric $SU(5)$ gGHU model 
compactified on an $S^1/\Z2$ orbifold, 
with the compactification scale $1/R$ 
being the GUT scale $\M{GUT}$.
The SUSY breaking scale $\M{SB}$ can be 
either the intermediate scale or a lower scale.
We start with the bulk symmetry 
$SU(5)_1\times SU(5)_2\times\Zex\times U(1)_D$,
where the $\Zex$ exchanges the two $SU(5)$ factors.
The bulk $SU(5)_1\times SU(5)_2$ symmetry is broken
by the orbifold boundary conditions (BCs)
down to its diagonal subgroup $SU(5)_V$, 
which is identified with the $SU(5)_G$.
This duplicated structure 
is a source of adjoint zero modes in the gGHU setup~\cite{gGHU1,gGHU2}.
Notice also that 
the bulk gauge group contains a $U(1)_D$ factor,
which is a basic ingredient for the Dirac gaugino models,
%
as will be reviewed in \red{Sect.}~\ref{Sec:DiracGaugino}.
%
Correspondingly
the CS-like term to be generated is related to a mixed anomaly 
between the $U(1)_D$ and $SU(5)$ gauge groups.
Therefore
we will refer to it as mixed CS-like term in the present paper.

This article is organized as follows. 
\red{%
In the following two sections, 
brief reviews are given respectively 
of the Dirac gaugino and the gGHU scenarios.
%
%
In Sect.~\ref{Sec:DiracGaugino},
we summarize the basic assumptions of 
the Dirac gaugino models.
We also comment on the issue of the lemon-twist (LT) operator
and its proposed solution~\cite{GoldstoneGaugino1,GoldstoneGaugino2}.
}
%
In Sect.~\ref{Sec:gGHU}, 
we review some elements of the $SU(5)$ gGHU models.
Specifically we explain how adjoint zero modes arise 
in a model with $SU(5)_1\times SU(5)_2\times\Zex$.
We also explain how to obtain incomplete GUT multiplets
in the gGHU setup.
%
In Sect.~\ref{Sec:CSingGHU}, 
we examine the $\Z2$ properties of the mixed CS-like term
and explain how required $\Z2$ breaking can be incorporated
in a consistent $S^1/\Z2$ orbifold.
We also derive a concrete expression for
the coefficient of the Dirac gaugino mass terms.
The section~\ref{Sec:summary} is devoted to summary and discussion.
%
%
\red{%
In Appendix~\ref{Sec:SUSY},
we summarize the field contents and
the supersymmetric Lagrangian of our model;
we also outline how
the supersymmetric CS-like term is related to 
the supersoft term responsible for the Dirac gaugino mass terms.
}
After a review on anomaly inflow on orbifold
and its relation to the CS term
%
\red{%
in Appendix~\ref{Sec:AnomalyInflow},
}
the detailed calculations of the mixed CS-like term are given 
in Appendix~\ref{Sec:calc} 
using a simplified setup.


%
\section{Elements of Dirac gaugino}
\label{Sec:DiracGaugino}
%


Here we give a brief review on models of Dirac gaugino,
recalling the basic assumptions behind the construction.
We start with a supersymmetric model that contains an adjoint chiral
superfield $\Phi^a$ 
for each gauge group $G_A$ in the SM ($A=3,2,1$),
\begin{align}
      \Phi^a\fun{x,\theta}
\ =\ 
     \phi^a\fun{x}+\sqrt{2}\theta^\alpha\psi^a_\alpha\fun{x}+\cdots
     \ ,
\label{Eq:adjoint}
\end{align}
where $\theta^\alpha$ is the superspace coordinate and
the adjoint index of $G_A$ is denoted by $a$.
We assume that supersymmetry is broken 
by a nonvanishing $D$-term $\VEV{D_D}$ of a hidden-sector $U(1)_D$.
Then the Dirac gaugino mass term can be obtained 
if we further assume that
integration of messenger sector fields gives rise to
the so-called supersoft operator~\cite{supersoft, Polchinski:1982an}
\bequ
 \mathcal{L}_{\mathrm{supersoft}}^{4d}
\  =\  \frac{C_Ag_A^2 g_D}{\Lambda}\int\!\mathrm{d}^2\theta\,
       \sqrt{2}\,\Phi^a W^a_\alpha W_D^\alpha
       \ .
\label{Eq:classic}
\eequ
Here 
$\Lambda$ is a mass scale at which the above operator is generated;
$W^a_\alpha$ ($W^D_\alpha$) are 
the field strength superfield of the SM gauge group $G_A$ 
(the hidden-sector $U(1)_D$ factor), respectively.
In \Eq{Eq:classic},
we have put the coefficient $C_A$
as well as the gauge coupling $g_A$ and $g_D$
of the gauge group $G_A$ and $U(1)_D$ respectively.
Substituting the nonvanishing $D$-term,
$\VEV{W_D^\alpha}=\theta^\alpha\VEV{D_D}$, into the supersoft operator,
we obtain a Dirac mass, 
\begin{align}
m_{D_A}
\ =\ 
     C_Ag_A^2\,
     \frac{g_D\VEV{D_D}}{\Lambda}
     \ ,
\label{Eq:gauginomass}
\end{align}
of the $G_A$-gaugino and the fermion component of $\Phi^a$.

An intriguing property of Dirac gaugino models comes from the fact that
the supersoft operator contains a trilinear coupling
of the scalar component $\phi^a$ of $\Phi^a$ to the other scalar fields.
This has two important consequences~\cite{supersoft}.
The first one is the supersoftness, that is, 
radiative corrections to a scalar mass become finite
since usual logarithmic divergences are canceled
by the adjoint scalar loop.
The second consequence is $D$-term cancellation, which means that 
the usual $D$-term contributions to 
the quartic scalar couplings
are canceled by tree-level exchange of the adjoint scalar 
%
\red{
$\phi^a$.
}


Let us summarize the assumptions in Dirac gaugino models:
(i) the presence of light adjoint chiral fields whose
fermion components are the Dirac partner of the gauginos,
(ii) the generation of the supersoft operator,
and (iii) $D$-term SUSY breaking in the hidden-sector.
As we see in Sect.~\ref{Sec:gGHU},
the assumption (i) can naturally be explained
in a supersymmetric version of the gGHU setup.  
To discuss the assumption (ii)
in such setup is the main purpose
of the present work
and is given in Sect.~\ref{Sec:CSingGHU}.
As for the SUSY breaking,
we just note that an example of dynamical SUSY breaking 
with a nonzero $D$-term is provided 
by the $SU(4)\times U(1)$ model in Ref.~\cite{Affleck:1984xz};
another example is Nambu--Jona-Lasinio type models 
of Refs.~\cite{Itoyama:2011zi,Itoyama:2012fk,D-termBreaking}.

%
We add a comment on masses of the adjoint scalars
$\phi^a=\left(\sigma^a+i\pi^a\right)/\sqrt{2}$.
The supersoft operator (\ref{Eq:classic})
gives a mass $2m_D$ to the real part $\sigma^a$,
but its pseudo-scalar partner $\pi^a$ 
remains massless \cite{supersoft}.
One expects radiative corrections to their masses,
except for the singlet. 
Phenomenologies with the chiral adjoints
are quite different depending on 
whether there is a superpotential coupling 
to the Higgs doublets.

%
%
\subsection{Problem of adjoint scalar mass \red{and its solutions}}
%
\label{Sec:AdjointScalarMass}
%


The successful generation of the supersoft term (\ref{Eq:classic})
is not the end of the story:
we should take care that unwanted terms are not generated
at the same time.
Among others, 
there is the so-called lemon-twist (LT) 
operator~\cite{Carpenter:2010as, Csaki:2013fla},
\bequ
 \mathcal{L}_{\mathrm{LT}}^{4d}
 = \frac{1}{\Lambda^2}\int\!\mathrm{d}^2\theta\,
   \Phi^a\Phi^a W_{D\alpha} W_D^\alpha \ , 
\label{Eq:LT}
\eequ
which contributes to a $B$-term like, 
holomorphic mass term of $\phi^a$. 
This contribution, if present,
decreases one eigenvalue of the mass squared 
of the scalar component $\phi^a$ 
to make it tachyonic.
This is problematic especially 
when both the operators in Eqs.~(\ref{Eq:classic}) 
and (\ref{Eq:LT}) are generated at one loop level.

A solution to this problem was proposed 
in Ref.~\cite{GoldstoneGaugino1,GoldstoneGaugino2}:
%
\red{
if
}
the scalar component $\mathrm{Im}\,\phi$ behaves 
as a (pseudo-)Goldstone field of a broken anomalous symmetry,
then the LT term (\ref{Eq:LT}) is forbidden 
by the shift symmetry of the Goldstone mode,
while the desired term (\ref{Eq:classic})
is still generated from the anomaly. 
Such scenario was called Goldstone gaugino scenario.


In this respect, it is interesting to note that 
extra-dimensional components of gauge fields 
are kinds of Nambu-Goldstone 
modes related to 
the breaking of 4D gauge symmetry on each point (4D slice) 
in the extra dimensions.
This fact is clear in the lattice regularization 
or deconstruction~\cite{deconstruction} of the extra dimensions.
As a result, 
if the adjoint chiral fields 
predicted in the gGHU setup 
are identified with the Dirac partner of the gauginos,
the resulting model may be regarded as
an extra-dimensional realization of the Goldstone gaugino scenario.
We note that in the gGHU picture,
the absence of the LT operator can be understood 
directly from the 5D gauge invariance.

\section{Review of grand gauge-Higgs unification}
\label{Sec:gGHU}
%

The basic idea of the gGHU scenario is 
to break the unified gauge symmetry by the Hosotani mechanism.
The gGHU, when applied to the SUSY $SU(5)$,
has intriguing properties:
a natural realization of the doublet-triplet (DT) splitting
and the prediction of light chiral adjoint fields.
Here we will explain these properties,
recalling some elements of orbifold construction for later purposes.

{}For definiteness,
let us consider a five-dimensional $SU(5)$ model 
with a simple Lagrangian
\bequ
 {\cal L}_5
  ={}-\frac14 F^a_{MN}F^{aMN}
     +\bar\Psi_{\bf R}(i\sla D-\gamma_5 D_5-m)\Psi_{\bf R} \ , 
\label{Eq:L5}
\eequ
where $F^a_{MN}$, $\Psi_{\bf R}$ and $D_M$ are respectively 
the field strength, 
a fermion field belonging to the {\bf R} representation
and the covariant derivative acting on it.
The 5D Lorentz and the adjoint indices are denoted
by $M=(\mu,5)=(0\mbox{-}3,5)$ and $a$ respectively.
[We can add further fermions, scalar fields and/or matter interactions.]
We will consider only the case without supersymmetry,
but it is straightforward to supersymmetrize the whole setup
by replacing the gauge (fermion) field 
with the vector (chiral) supermultiplet.

To realize the chiral fermions of the SM, 
we compactify the fifth dimension on an $S^1/\Z2$ orbifold, 
which is a quotient space of a circle $S^1$, 
divided by the identification under the 5D parity $P_5: x^5\to-x^5$. 
Two fixed points are denoted by $x^5_p\equiv p\pi{R}$ ($p=0,1$).
The circle with the radius $R$ 
can be regarded as a quotient of the covering space $\mathbb R$ 
divided by the translation $T: x^5\to x^5+2\pi R$, and 
the product $P'_5=TP_5$ generates the parity around $x^5=x^5_1$; 
that is, $P'_5: \pi R-x^5\to\pi R+x^5$.

When the theory has another $\Z2$ symmetry, 
the identification can be twisted;
for instance if we choose a nontrivial element of the gauge group, 
$P_g$, as the generator of the additional $\Z2$, 
the bulk gauge symmetry can be reduced 
by the orbifold BCs~\cite{%
orbifoldGUTs1,orbifoldGUTs2,orbifoldGUTs3,%
orbifoldGUTs4,orbifoldGUTs5,orbifoldGUTs6%
}. 
This can be understood by applying the identification 
to the fields~\footnote{
Here $P_g$ is an abstract group element while 
$\widehat{P}_g$ is the corresponding matrix 
in the defining representation.
}
\begin{align}
 \Lm A_M(x^5),\,\Psi_{\bf R}(x^5)\Rm
 &= P_5P_g \Lm A_M(x^5),\,\Psi_{\bf R}(x^5)\Rm
 \nn\\
 &=\Lm (-)^M\widehat{P}_gA_M(-x^5)\widehat{P}_g^\dagger,
   \,\eta_\Psi\gamma_5\rho_{\bf R}[P_g]\Psi_{\bf R}(-x^5)\Rm,
\end{align}
where $(-)^M$ takes $+1(-1)$ for $M=\mu\,(5)$, 
and $\rho_{\bf R}[P_g]$ denotes the matrix representation of $P_g$ on the fermion $\Psi_{\bf R}$.
A parallel discussion holds for $P'_5$ 
with (generally different) $P'_g$.
Note that 
for each fermion $\Psi$, 
the sign factor $\eta_\Psi$ can be $+1$ or $-1$.
A similar sign degrees of freedom 
$\eta_\Psi'$ and $\eta_T\left(=\eta_\Psi\eta_\Psi'\right)$
exist for the parity $P'_5$ and the translation $T$ respectively.
On the other hand,
there is no such sign degree of freedom for the gauge field. 
The components of $A_\mu$ that commute with $P_g$ and $P'_g$
have a zero mode and correspond to the low-energy gauge symmetry;
the other components, not even functions of $x^5$ or $x^{5\prime}=x^5-\pi R$,
do not have zero modes and thus
decouple from the low-energy theory.
Similarly, the components of $A_5$ that 
\textit{anti}-commute with both $P_g$ and $P'_g$ have zero modes.
In the usual Hosotani mechanism,
it is these zero modes that acquire a nontrivial VEV
to break the gauge symmetry further.

%
\subsection{Adjoint zero modes via diagonal embedding}
\label{Sec:adjoint}
%

To apply the Hosotani mechanism to the $SU(5)_G$  breaking, 
we need an adjoint zero mode of $A_5^a$.
Due to the factor $(-)^M$, however, 
$A_5^a$ has the 5D parity opposite to $A_\mu^a$,
and so the $A_5^a$ does not have the zero mode
in the (adjoint) component 
corresponding to the zero mode gauge fields $A_\mu^a$
of the unbroken gauge group.
Actually a way of realizing the adjoint zero modes
is provided by
the diagonal embedding method, which was developed 
in the context of the string theory~\cite{%
DiagonalEmbedding1,DiagonalEmbedding2,DiagonalEmbedding3,%
DiagonalEmbedding4,DiagonalEmbedding5,DiagonalEmbedding6,%
DiagonalEmbedding7,DiagonalEmbedding8%
}
and applied to our field theoretical setup~\cite{gGHU1,gGHU2}.

For this purpose, we introduce two copies of the gauge group 
and suppose that
there is a $\Zex$ symmetry that exchanges the two gauge fields,
$A\supP{1}_M$ and $A\supP{2}_M$.
We denote by $P_{ex}$ the generator of this $\Zex$,
\bequ
P_{ex}:\  A\supP{1}_M(x)\ \longleftrightarrow A\supP{2}_M(x) \ .
\eequ
Then the orbifold BCs that give rise to the desired adjoint zero modes
are given by the combined (or simultaneous) action of 
the 5D parity $P_5$ 
and the $\Zex$ exchange $P_{ex}$:
\bequ
     \begin{pmatrix}A\supP{1}_M(x^5) \,,& A\supP{2}_M(x^5)\end{pmatrix}
 =
     (-)^M  
     \begin{pmatrix}A\supP{2}_M(-x^5) \,,& A\supP{1}_M(-x^5)\end{pmatrix}
\label{Eq:BC:gauge}
\eequ
around the first fixed point $x^5_0=0$, 
and similar ones around the other fixed point $x^5_1=\pi{}R$.
%
%
Defining the $\Zex$ eigenstates 
by $X\supP{\pm}\equiv(X\supP1 \pm X\supP2)/\sqrt2$,
we see that
$A\supP+_\mu$ and $A\supP-_5$ have zero modes.
This means that
the gauge symmetry is reduced to the diagonal subgroup of the two gauge groups
and that the $A\supP-_5$ zero mode behaves 
as an adjoint field under the remaining gauge symmetry.
In this way, 
we obtain the adjoint scalar field that can be used to break
the diagonal subgroup further.

Notice that, in this type of GHU scenario, 
the Higgs field which is unified with the gauge field is 
not the SM Higgs field, but the adjoint Higgs field 
that breaks the GUT gauge symmetry down to the SM one.
After the $SU(5)_G$ breaking, 
a part of adjoint fields are absorbed via the Higgs mechanism, 
and more importantly,
\textit{%
there appear the adjoint scalar fields of the SM gauge group,
}%
namely, the color octet, the weak triplet and the singlet fields,
in the low-energy effective theory.
As for the SM Higgs, we need a separate consideration 
as we review shortly.


As for fermions,
we introduce a $\Zex$ pair of bulk fermions:
$\Psi_{({\bf R_1},{\bf R_2})}\supP{1}$ 
belonging to ${\bf R_1}$ (${\bf R_2}$) 
representation of the first (second) gauge group,
and its $\Zex$ partner $\Psi_{({\bf R_2},{\bf R_1})}\supP{2}$.
The BCs for them are given by
%
\bequ
     \begin{pmatrix}
       \Psi_{({\bf R_1},{\bf R_2})}\supP{1}(x^5)\,,&
       \Psi_{({\bf R_2},{\bf R_1})}\supP{2}(x^5)
     \end{pmatrix}
 =
     \eta_\Psi\gamma_5 
     \begin{pmatrix}
       \Psi_{({\bf R_2},{\bf R_1})}\supP{2}(-x^5) \,,&
       \Psi_{({\bf R_1},{\bf R_2})}\supP{1}(-x^5)
     \end{pmatrix}
     \ ,
\label{Eq:BC:psi}
\eequ
%
and similar ones for $P'_5$.
We summarize in Table~\ref{Table:Z2}
the parity eigenvalues of each field under the $P_5P_{ex}$ and $P_5'P_{ex}$.
Note that
the BCs of $\Psi\supP{\pm}_L$ are the same as those of
$\Psi\supP{\mp}_R$ with the opposite $P_{ex}$ parity. 
Consequently, for instance,
when $\Psi\supP{+}_L$ has the zero mode, $\Psi\supP{-}_R$ also does. 
Hereafter we set ${\bf R}_2={\bf 1}$ for simplicity.
In this case,
these fields $\Psi\supP{+}_L$ and $\Psi\supP{-}_R$ belong to 
the same representation under the remaining gauge symmetry.
Therefore a bulk fermion in the present setup gives rise to
zero modes in a vector-like representation with the opposite
$P_{ex}$ parity.
The chiral fermions, such as the SM fermions, may be put on the boundaries.

\begin{table}[tbp]
\caption{%
The boundary conditions: 
the first (second) sign shows the orbifold parity 
of each field around the boundary at 
$x^5_0=0$ ($x^5_1=\pi R$).
}
\label{Table:Z2}
 \begin{center}
  \begin{tabular}{cccclcccc} 
    \hline
    $A_\mu\supP+$ & $A_\mu\supP-$ & $A_5\supP+$ & $A_5\supP-$ & \quad & 
    $\Psi_R\supP+$ & $\Psi_R\supP-$ & $\Psi_L\supP+$ & $\Psi_L\supP-$
    \rule[0mm]{0mm}{5mm} \\[0mm]\hline\hline
    $(+, +)$ & $(-, -)$ & $(-, -)$ & $(+, +)$ && 
    $(\eta_\Psi, \eta'_\Psi)$ & $(-\eta_\Psi, -\eta'_\Psi)$ & 
    $(-\eta_\Psi, -\eta'_\Psi)$ & $(\eta_\Psi, \eta'_\Psi)$
    \rule[0mm]{0mm}{4mm} \\[1mm]\hline
  \end{tabular} 
 \end{center}
\end{table}

%
\subsection{Doublet-triplet splitting and gauge coupling unification}
\label{Sec:DTS&GCU}
%

A striking feature of the present gGHU scenario is that 
the DT splitting can be naturally realized 
even in an $SU(5)$ model~\cite{gGHU-DTS}. 
This is possible on a specific vacuum where the Wilson line $W$, 
the order parameter of the $SU(5)$ breaking, is given, 
in the defining representation, 
by~\footnote{%
Alternatively we can gauge away the Wilson line $W$, so that
the BC at $x^5_1$ 
is modified to 
$P'_g W$.
}
\bequ
 W={\cal P}\exp\Ls i\frac{g}{\sqrt{2}}\int A_5^{(-)} dx^5\Rs=\diag(1,1,1,-1,-1) \ , 
\label{WilsonLine}
\eequ
where 
$\cal P$ denotes the path-ordering
and $g$ is the 5D gauge coupling constant. 
Since the unity in $W$ corresponds to the trivial vacuum 
with vanishing $\VEV{A_5}$, 
the above $W$ corresponds to 
the (inversely) missing VEV~\cite{DW1,DW2,DW3,DW4,DW5,DW6,DW7};
schematically, it is $\VEV{A_5}=\diag(0,0,0,v,v)$
with $v\neq0$.
Notice that usually 
the traceless condition of the $SU(5)$ generators
forbids this type of missing VEV;
in the present case, it is allowed 
since $W$ is an element of the $SU(5)$ group, not the algebra.

%
\red{%
{}For later purpose, 
let us introduce some notation for a diagonal $W$:
we write its diagonal component as $w=e^{ai\pi}$.
Consequently,
we can express the $SU(5)_G$-breaking VEV (\ref{WilsonLine})
by stating that 
$e^{ai\pi}=+1$ for the $SU(3)$ and
$e^{ai\pi}=-1$ for the $SU(2)$ subgroups.

}

%
%

The Wilson line (\ref{WilsonLine}) 
can be used to realize the DT splitting.
To see this, we introduce a pair of the bulk Higgs fields,
$H_{(\bf{5},\bf{1})}\supP{1}$ and $H_{(\bf{1},\bf{5})}\supP{2}$,
in the $SU(5)_1 \times SU(5)_2$ setup.
Na\"{i}vely the missing VEV contributes to 
the doublet mass instead of the triplet mass.
Instead, assigning
an antiperiodic BC to the bulk Higgs multiplets,
namely, $\eta_T=\eta_H\eta_H'=-1$, 
we obtain the zero mode only in the doublet component. 
In this way we can naturally realize
the correct pattern of $SU(5)$ gauge symmetry breaking
and the DT splitting
if the expectation value of the Wilson line is given
by \Eq{WilsonLine}.

We note that
the required value (\ref{WilsonLine}) respects 
the $\Zex$ exchange symmetry of $SU(5)_1\times SU(5)_2$:
the $P_{ex}$ transformation flips the sign of $A_5\supP-$, 
which transforms $W$ to its complex conjugate $W^*$. 
This will guarantee that the vacuum is a extremum of 
dynamically generated effective potential without fine-tuning,
although we will not attempt to analyze it here.

Another remark concerns the gauge coupling unification,
which is realized in the minimal SUSY SM 
but is ruined by the adjoint chiral supermultiplets
and/or by a deviation of 
the SUSY-breaking scale $\M{SB}$ from TeV scale.
In this respect, note that
the above mechanism for the DT splitting can also 
generates a mass splitting in a bulk hypermultiplet 
other than the $(\bf{5},\bf{1})$ and $(\bf{1},\bf{5})$ Higgs fields.
This allows us to introduce 
(vector-like) $SU(5)$ incomplete multiplets in the 4D effective theory. 
For instance,
a $\Z2^{ex}$ pair of  periodic 
${\bf10}$ superfields give vector-like pairs
whose quantum numbers of the SM gauge group are the same 
as the right-handed up quark and the right-handed charged lepton 
respectively. 
The gauge coupling unification can be recovered 
by a suitable choice of additional bulk superfields.
A concrete example was given in Ref.~\cite{gGHU-DTS}.



%
\section{Chern-Simons term in grand gauge-Higgs unification}
\label{Sec:CSingGHU}
%

As we reviewed in the previous section, the grand GHU model, 
a supersymmetric $SU(5)_1\times SU(5)_2$ theory on $S^1/\Z2$ orbifold,
is a natural starting point for constructing models of Dirac gaugino.
It ``predicts'' the light chiral adjoint $\Phi^a$,
a chiral supermultiplets (\ref{Eq:adjoint})
in the adjoint representation of the SM gauge group.
%
\red{%
We can identify its fermionic component
as a Dirac partner of each gaugino.
}
%
Then the next task is to generate the supersoft operator 
(\ref{Eq:classic}) in the gGHU setup.
For this purpose,
we will be interested in a particular bosonic term,
$\phi^a F^a_{\mu\nu} \widetilde{F}^{\mu\nu}_D$,
contained in that operator.
A straightforward supersymmetrization~\cite{HigherDimSUSY}
 will lead to the desired operator,
as we sketch in Appendix~\ref{Sec:Matching}.

We consider a 5D $SU(5)_1\times SU(5)_2\times U(1)_D$ gauge theory,
extending the gGHU to include the $U(1)_D$ gauge group 
responsible for the mediation of SUSY breaking;
we assume that the $U(1)_D$ gauge field is $\Zex$-even
so that it has a zero mode,
and denote its field strength by $F^D_{\mu\nu}$.
Notice that
the supersoft operator (\ref{Eq:classic}) contains 
a term of the form 
$\phi^a F^a_{\mu\nu} \widetilde{F}^{\mu\nu}_D$,
where 
$\phi^a=\left(\sigma^a+i\pi^a\right)/\sqrt{2}$
is the adjoint scalar.
In the gGHU setup,
the pseudo-scalar component $\pi^a$ 
arises from the fifth component of the gauge field, $A_5^{(-)}$,
while the 4D gauge fields are 
from the zero modes of $A_\mu^{(+)}$.
In this way,
we are led to the idea that
the 5D counterpart of the supersoft operator (\ref{Eq:classic}) 
in the gGHU scenario is given by a mixed CS-like term
\bequ
     A^{a(-)}_5 F^{a(+)}_{\mu\nu}\widetilde{F}^{\mu\nu}_D
\ =\ \frac{1}{2}\,\epsilon^{\mu\nu\rho\sigma}
     A^{a(-)}_5 F^{a(+)}_{\mu\nu}{F}^D_{\rho\sigma}
     \ .
\label{Eq:CS:goal}
\eequ
To discuss how this term can be generated,
we define the CS-like term of $SU(5)_i^2\times U(1)_D$ by
\bequ
\mathcal{O}^{(i)}
\ =\ 
     {A^{a(i)}_5}{F^{a(i)}_{\mu\nu}} \widetilde{F}^{\mu\nu}_D
     \ ,
\label{Eq:CS:def}
\eequ
where $i=1,2$.
%
The desired \red{operator}
%
(\ref{Eq:CS:goal}) is contained in a combination
\bequ
\mathcal{O}^{(-)}
\ =\ \mathcal{O}^{(1)}-\mathcal{O}^{(2)}
\ =\ A_5^{a(-)} F_{\mu\nu}^{a(+)}\widetilde{F}_D^{\mu\nu}
    +A_5^{a(+)} F_{\mu\nu}^{a(-)}\widetilde{F}_D^{\mu\nu}
     \ ,
\label{Eq:CSingGHU}
\eequ
where
the last term on the right-hand side contains at least one massive KK mode 
and thus decouples from the low-energy effective theory. 
We see from the relation (\ref{Eq:CSingGHU}) that 
the generation of the 
$\mathcal{O}^{(-)}=\mathcal{O}^{(1)}-\mathcal{O}^{(2)}$ 
requires asymmetry between $SU(5)_1$ and $SU(5)_2$,
that is, a sort of breaking effect of the $\Zex$ invariance.

Now, it is important to notice that
%
\red{the} desired \red{operator}
%
(\ref{Eq:CS:goal}) is odd under the 5D parity;
it is also odd under the $\Zex$.
To generate such term,
we should break the $\Zex$ invariance as well as the $\Zp$ parity
of the bulk Lagrangian.

One may wonder whether
the breaking of the $\Zp=\left\{1,P_5\right\}$ 
or the $\Zex=\left\{1,P_{ex}\right\}$ 
would be unacceptable for a consistent construction of orbifold.
Actually it is not the full $\Zp\times\Zex$ invariance
but its diagonal subgroup $\Zcomb = \left\{1,P_5P_{ex}\right\}$
that is required from the consistency of the theory.
In other words, we require only the invariance 
under the simultaneous action of the $P_5$ and the $P_{ex}$.

%
\subsection{Model\red{: fermion sector}}
%
\label{Sec:model}
%

With the above in mind, let us present our concrete setup 
for generating the CS-like term (\ref{Eq:CSingGHU}),
through anomaly inflow mechanism sketched 
in Appendix~\ref{Sec:AnomalyInflow}.
%
\red{%
In discussing such anomaly inflow,
we can focus on the gauge-fermion sector of the model,
whose supersymmetric form is summarized in Appendix~\ref{Sec:SUSY}.
}


We introduce 
%
\red{%
a $\Z2^{ex}$ pair of bulk hypermultiplets,
whose fermionic components can be represented by
a 5D Dirac fermion 
}
%
$\Psi_{({\bf R},{\bf 1})}\supP{1}$ 
and its $\Z2^{ex}$ partner $\Psi_{({\bf 1},{\bf R})}\supP{2}$.
%
\red{%
Here $\mathbf{R}$ is an $SU(5)$ representation 
such as $\mathbf{\bar{5}}$, $\bf{10}$.
We refer to them as messenger multiplet.
}
These \red{messenger} fermions contain zero modes 
in $\Psi_L\supP{+}$ and $\Psi_R\supP{-}$.
We also add the boundary-localized field $\psi_L^p$ 
on each boundary at $x^5=x^5_p$ 
that belongs to 
%
\red{
the representation ${\bf R}$ of the diagonal $SU(5)$.
}
The $U(1)_D$ charges of these fermions are commonly set to $q_D$. 
%
\red{%
The field contents in the fermion sector 
is summarized in Table~\ref{tab:content:su5},
where quantum numbers of the boundary-localized fields
are shown only for the unbroken symmetries.
}

\begin{table}[tb]
\caption{
	Field contents of the fermion sector
        relevant for calculating the mixed CS-like term:
\red{%
        see Table~\ref{tab:content:susy} 
        for the supersymmetric setup.
}%
}
\begin{center}
\begin{tabular}{c||cc|c|c}
\hline
	\makebox[12mm]{bulk fields} & 
	\makebox[30mm]{	$SU(5)_1{\times}SU(5)_2$ } & 
	\makebox[15mm]{	$SU(5)_G$ } & 
	\makebox[10mm]{ $U(1)_D$ } & 	
	\makebox[10mm]{	$P_5$ } 
	\\\hline 
	$\Psi_L^{(1)}(x^M)$ &
	$(\,\mathbf{R},\mathbf{1}\,)$ &	$\mathbf{R}$ & 
        $q_D$ & $+$ 
	\\
	$\Psi_L^{(2)}(x^M)$ &
	$(\,\mathbf{1},\mathbf{R}\,)$ &	$\mathbf{R}$ & 
        $q_D$ & $+$ 
	\\
	$\Psi_R^{(1)}(x^M)$ &
	$(\,\mathbf{R},\mathbf{1}\,)$ &	$\mathbf{R}$ & 
        $q_D$ & $-$ 
	\\
	$\Psi_R^{(2)}(x^M)$ &
	$(\,\mathbf{1},\mathbf{R}\,)$ &	$\mathbf{R}$ & 
        $q_D$ & $-$ 
        \\
	\hline \hline 
	boundary fields &
	localized position & 
	\makebox[15mm]{	$SU(5)_G$ } & 
	\makebox[10mm]{ $U(1)_D$} & 	
	\makebox[10mm]{	$P_{ex}$ } 
        \\ \hline 
 $\psi_L^{p=0}(x^\mu)$ & 
        $x^5_0=\ 0\ $  & $\mathbf{R}$ & $q_D$ & $-$
 \\ 
 $\psi_L^{p=1}(x^\mu)$ & 
        $x^5_1=\pi{}R$ & $\mathbf{R}$ & $q_D$ & $-$
 \\\hline
\end{tabular}
\end{center}
\label{tab:content:su5}
\end{table}


The 5D Lagrangian for these fermions is given by 
$\mathcal{L}_\Psi
=\mathcal{L}_\Psi^{\mathrm{bulk}}
+\mathcal{L}_\Psi^{\mathrm{boundary}}$, 
where\,\footnote{
Our Lagrangian is normalized when integrated over the interval
$\left[0,\,\pi R\right]$.
Therefore in \Eq{Eq:Lboundary},
we put a factor of $2$ in front of the delta functions.
}
\beqn
\mathcal{L}_\Psi^{\mathrm{bulk}}
   &=& \sum_{i=1,2}
     \bar\Psi\supP{i}(i\sla D-\gamma_5D_5-m_i)\Psi\supP{i}
     \ , 
\label{Eq:Lbulk}\\
\mathcal{L}_\Psi^{\mathrm{boundary}}
   &=& \sum_{p=0,1}
     \left\{
       \bar\psi_{L}^pi\sla D\,\psi_{L}^p
      -\sqrt{\mu_p}\left(\bar\psi_{L}^p\Psi_R\supP-+c.c.\right) 
    \right\} 2\delta\!\left(x^5-x^5_p\right) 
     \ .
\label{Eq:Lboundary}
\eeqn
Here
$D_M$ is the covariant derivative with respect to
$SU(5)_1 \times SU(5)_2 \times U
(1)_D$,
and $m_{i=1,2}$ are bulk mass parameters.
With a choice $m_1=-m_2=m_-$,
the bulk fermion mass terms take the form
\beqn
{}-{\cal L}_{\mathrm{mass}}^{\mathrm{bulk}}
\ =\ m_-\left\{
        \bar\Psi\supP{1}\Psi\supP{1}-\bar\Psi\supP{2}\Psi\supP{2}
        \right\}
\ =\ m_-\left\{
        \bar\Psi\supP+\Psi\supP-+\bar\Psi\supP-\Psi\supP+
        \right\}
 \ .
\label{Eq:Lbulkmass}
\eeqn
For notational simplicity we will denote $m_-$ just by $m$,
except in Appendix~\ref{Sec:calc} where
a common piece $m_+$ of the bulk mass parameters is also added.

Some remarks are in order.
Our choice of the bulk mass parameters, $m_1=-m_2=m$,
explicitly breaks the $\Zex$ invariance.
Note that we are considering a mass term 
that is constant in the fifth dimension
instead of the ``usual'' kink mass term.
Consequently
the bulk messenger mass term (\ref{Eq:Lbulkmass})
breaks the $\Z2^{5d}$ as well as the $\Zex$
while keeping the simultaneous $\Zcomb$, that is,
\begin{align}
\Z2^{5d} \times \Z2^{ex}
\  \longrightarrow\ \Z2^{comb} \ .
\label{Eq:Z2breaking}
\end{align}
Therefore
such a bulk mass term 
is allowed in a consistent orbifold construction.
We assume that
\textit{%
the bulk mass term for the messenger fields
is the unique source of the $\Z2$-breaking (\ref{Eq:Z2breaking})},
so that
the bulk mass parameter $m$ characterizes its breaking effects.
We expect that the required CS-like term, 
if generated successfully, 
is proportional to a power of this mass parameter.
We will take $m$ as a free parameter.

The second remark concerns fermion zero modes.
In the absence of our bulk mass term,
the messenger fermions contain zero modes 
in $\Psi_L\supP{+}$ and $\Psi_R\supP{-}$ components.
Once we add the bulk mass term, however,
the would-be massless modes 
will acquire a mass 
as can be seen from the second form
in Eq.~(\ref{Eq:Lbulkmass}).

The absence of the fermion zero modes will complicate 
the following discussion of generating the CS-like term.
One can still calculate the effective action 
by integrating out the heavy messenger fields, 
which we will not do here in the present paper.
Instead, we focus on the possibility of
determining the CS-like term
through chiral anomaly induced by fermion zero modes.
This can be achieved 
by introducing boundary fields, $\psi^p_L$,
that have bulk-boundary mixing mass terms 
with one of the zero modes, $\Psi_R\supP-$,
as in Eq.~(\ref{Eq:Lboundary}).
The total anomaly can be canceled 
by introducing additional fields, 
as we shall discuss shortly in Sect.~\ref{Sec:anomaly}.

We note that 
for the existence of a fermion zero mode,
it is enough to add a single boundary fermion,
$\psi_L^0$ at $x^5=0$ or $\psi_L^1$ at $x^5=\pi R$,
but we consider adding both for definiteness.

%
\subsection{Mixed CS-like term and Dirac gaugino mass}
\label{Sec:result}
%

Given the Lagrangian as above, 
we analyze the bulk equations of motion (EOMs) 
and boundary conditions 
to find the KK spectrum and wavefunctions.
In the present paper,
we confine ourselves to the limiting case
where the bulk-boundary mixing masses are very large,
$\mu_p\gg{}m$ and $\mu_p\gg{}1/R$.
In this case, the two zero modes are dominantly contained 
in $\Psi_L\supP1$ and $\Psi_L\supP2$, 
\begin{eqnarray}
\Psi_L\supP{i}(x^\mu,x^5)
\ =\ \psi_{L0}\supP{i}(x)\,\xi_{L0}\supP{i}(x^5) +\cdots
     \ , \qquad
 \xi_{L0}\supP{i}(x^5) \propto\ e^{m_i x^5} 
     \ ,
\label{Eq:bulkzeromode}
\end{eqnarray}
which are localized, 
by our choice $m_1=-m_2=m$,
to the opposite boundaries.

Using the profiles (\ref{Eq:bulkzeromode}) of the zero modes, 
we can calculate the coefficient of the effective mixed CS-like term.
We defer detailed calculations to Appendix~\ref{Sec:calc}, 
using a simplified setup with 
$SU(5)$ gauge groups replaced by $U(1)$'s.
The result for $U(1)_1\times U(1)_2\times U(1)_D$ model
is given by \Eq{Eq:4D-CSterm:U1}.
%
With a straightforward modification of group-theoretical factors,
the result for the $SU(5)_1\times SU(5)_2\times U(1)_D$ model
is given by
\bequ
\mathcal{L}_{\mathrm{CS}}^{4d}
\ =\ 
    \frac{2\,T(\mathbf{R}) g_G^2\,{q_D} g_D }{16\pi^2}\,
    {\pi R}
    f\!\left(\modd\pi R\right)
    \,{A^{a(-)}_5}{F_{\mu\nu}^{a(+)}}\widetilde{F}^{\mu\nu}_D
    \ , 
\qquad
f(z)=\frac{1}{\tanh{z}}-\frac{1}{z} \ ,
\label{Eq:4D-CSterm:SU5}
\eequ
where 
$g_G$ and $g_D$ are the 4D gauge coupling constants of 
$SU(5)_G\times U(1)_D$,
and $q_D$ is the $U(1)_D$ charge of the messenger fermion fields.
The group-theoretical factor $T({\bf R})$ is defined by 
${\rm tr}(t_{\bf R}^at_{\bf R}^b)=T({\bf R})\delta^{ab}$, 
using the $SU(5)$ generators $t_{\bf R}^a$ 
in the representation ${\bf R}$.

The dependence on the bulk mass parameter $\modd$
is contained in the function $f(z)$,
which is approximated by 
$f(\modd\pi R)\sim (\modd\pi R)/3$ for a small $\modd$,
while it approaches $1$ for a large $\modd$. 
Therefore the coefficient is typically of order of 
the inverse of the compactification scale, 
while it is suppressed for a small $\modd$, as is expected. 
Thus we conclude that 
the mixed CS-like term responsible for the Dirac gaugino mass 
is actually generated in our setup of a supersymmetric gGHU model.


Now, we can match the above result to the supersoft operator 
(\ref{Eq:classic}) with a coefficient $C_A$ for the gauge group $G_A$.
Identifying the mass scale $\Lambda$ in \Eq{Eq:classic}
with the compactification scale $\M{GUT}=1/R$,
we find that 
the coefficient of the supersoft operator
is given by 
\begin{align}
C_A
\ =\ \frac{2T(\mathbf{R})q_D }{16\pi}\,f\fun{m\pi R} 
     \ .
\end{align}
We see the coefficients are universal for the SM gauge group:
$A=3$ for $SU(3)$, $A=2$ for $SU(2)$ and $A=1$ for $U(1)$.
This feature is specific to the present limiting case of
large bulk-boundary mixing, 
$\mu_p\gg{}\modd$ and $\mu_p\gg{}1/R$,
where the coefficients become independent of
the Wilson line $a$:
%
%
$e^{a\pi i}=+1$ for the $SU(3)$ and 
$e^{a\pi i}=-1$ for the $SU(2)$.

In passing, we give an order estimate of
the resulting Dirac gaugino mass scale (\ref{Eq:gauginomass}).
For a rough estimate, we set gauge couplings to $\order{1}$.
Taking $\M{GUT}=1/R=\order{10^{16}\,\mathrm{GeV}}$
and $\sqrt{\VEV{D_D}}=\order{10^{12}\,\mathrm{GeV}}$
as a reference value,
we have 
\begin{align}
m_D
\ \approx\ 10^3\,\mathrm{GeV}\times
        \left(\frac{\sqrt{\VEV{D_D}}}{10^{12}\,\mathrm{GeV}}\right)^2
        \left(\frac{m}{10^{12}\,\mathrm{GeV}}\right)
\end{align}
for $m\ll \M{GUT}$.
Thus the gaugino mass of TeV scale is possible
for a moderate choice of parameters.
On the other hand, it has the upper bound
for a fixed value of SUSY-breaking VEV $\VEV{D_D}$,
since the dependence of the bulk mass parameter is
saturated for $m \gsim \M{GUT}$.
Therefore $m_D$ can be of the intermediate scale
only for a sufficiently large value of $\VEV{D_D}$:
for instance, $m_D=\order{10^{11}\,\mathrm{GeV}}$ 
for $\sqrt{\VEV{D_D}}=\order{10^{14}\,\mathrm{GeV}}$.


%
\subsection{Comments on other anomalies and CS-like terms}
\label{Sec:anomaly}
%

Up to now, we have focused on 
the mixed CS-like term (\ref{Eq:CS:goal}),
induced from the mixed anomaly of 
$SU(5)\supP-SU(5)\supP+U(1)_D$ spread in the bulk.
Here we make some comments on other anomalies and CS-like terms.

In the above setup,
other anomalies do not vanish even in the 4D effective theory,
including the cubic ${SU(5)\supP+}$ anomaly in particular.
As usual, we choose the matter content so that 
the 4D effective gauge symmetries are anomaly free. 
We can always cancel the $U(1)_D$ anomalies
by adding $SU(5)$-singlets, or,
by a suitable choice of the $U(1)_D$ charges.
For the cancellation of the $SU(5)$ anomaly, 
we can introduce the ``vector-like'' partners, 
$\psi_{R}^p(\mathbf{R},-,q_D)$, for the (chiral) boundary fermions.
In this case, however, we have to assume that
these partners have only small mixing to remaining fields,
not to disturb the above discussions. 
A more radical, interesting possibility is 
to identify the boundary fields as the SM matter fields:
Namely, we introduce the boundary fields
in the $\bf{\bar{5}}$ and $\bf{10}$ representations.
The representation $\bf{R}$ of the messenger multiplets
can be either $\bf{\bar{5}}$ and $\bf{10}$.
The mixing between bulk and boundary fields could 
play some roles for generating the structure of the Yukawa matrices.
We leave this possibility as a future work.

In any case we can choose a set of boundary matter fields
so as to cancel the cubic $SU(5)\supP{+}$ anomaly.
Even after the total anomaly is canceled, 
the corresponding CS-like term might be generated
so as to cancel the anomaly in the bulk,
in a similar manner in which 
the desired term (\ref{Eq:CS:goal}) is generated.
However, 
such CS-like term, even if generated, 
decouples from the 4D effective theory 
since it involves at least one field that has no zero mode.

Meanwhile, 
there is a CS-like term that does not decouple from the low-energy,
\begin{align}
\mathrm{tr}\left[
  A_5^{(-)}F_{\mu\nu}^{(+)}\widetilde{F}^{(+)\mu\nu}
\right] \ .
\label{Eq:axion-like}
\end{align}
This term will be generated
even if there is no net anomaly for $SU(5)\supP-{SU(5)\supP+}^2$.
Generically we expect that
its effects to the low-energy effective theory will be small
as it is a higher-dimensional operator.
(as far as the $D$-term of $SU(5)\supP+$ does not have a large VEV).
There is a possible exception, however.
%
Recall from \red{Sect.}~\ref{Sec:DiracGaugino} that 
%
the pseudo-scalar component of the adjoint chiral multiplet
do not get a mass from the SUSY breaking.
In particular the hypercharge component of $A_5^{(-)}$,
being a singlet under the SM gauge group,
remains massless.\footnote{
This is true if the adjoint chiral multiplets
have no superpotential coupling to the Higgs fields.
}
Since the above term (\ref{Eq:axion-like})
contains axion-like couplings 
of the hypercharge component of $A_5^{(-)}$
to the $SU(3)\times SU(2)\times U(1)$ gauge field strengths
in the SM,
it would be interesting to examine whether
such a component of $A_5^{(-)}$ 
can play a role of an axion-like field.



%
\section{Summary and Discussion}
\label{Sec:summary}
%

In this article, we have shown that
the grand gauge-Higgs unification model
is a good starting point for constructing the Dirac gaugino models;
the light adjoint chiral superfields predicted in the gGHU 
play a role of the Dirac partner of the gauginos,
and the supersoft term (\ref{Eq:classic}) 
can be obtained as a sort of the supersymmetric CS term 
in the 5D setup.
We have presented a concrete setup of field contents
and calculated the coefficient of the mixed CS-like term 
from the profile of chiral fermion zero modes.
The same result (and some generalization) can be obtained
by summing up the massive fermion KK modes,
as we shall show in a separate publication.

Our model may be regarded 
as an extra-dimensional realization 
of the Goldstone gaugino scenario,
proposed before as a solution to the adjoint scalar mass problem
in a generic model of Dirac gauginos.
In our present approach,
the absence of the LT operator (\ref{Eq:LT}) follows 
directly from the 5D gauge invariance.
We also note that
our model based on the supersymmetric gGHU 
supply 
a natural GUT completion of the Goldstone gaugino scenario.

A nontrivial point in our construction is 
the $\Z2^{5d} \times \Z2^{ex}$ properties 
of the desired CS-like term:
it is not invariant under the 5D parity $P_5$  nor 
the exchange $P_{ex}$ of the two $SU(5)$ groups.
To incorporate such $\Z2$ breaking effect
within a consistent orbifold compactification,
we introduce the bulk fermion mass term (\ref{Eq:Lbulkmass})
that is $P_5$-odd and $P_{ex}$-odd 
while invariant under the simultaneous action of $P_5$ and $P_{ex}$.
Consequently
the calculated coefficient of the mixed CS-like term, 
a function of the bulk mass parameter $\modd$,
vanishes in the limit $\modd\to0$,
and can be parametrically small
since it expresses the explicit breaking 
(\ref{Eq:Z2breaking}) of the $\Z2$ symmetries.

Specifically
we calculated the coefficient 
though anomaly inflow induced by fermion zero modes.
Since the $\Z2$ breaking 
by the bulk fermion mass term 
removes the (would-be) fermion zero modes,
we put boundary-localized fermions 
with bulk-boundary mixing masses $\mu_{p=0,1}$.
The boundary fields have an effect
of changing the boundary conditions for the bulk fields
especially when the mixing masses are large.
Interestingly, the obtained CS-like term becomes 
independent of the Wilson line $a$ in this limit.
The reason for the $a$-independence is that
the fermion zero modes are dominantly contained in the bulk fields, 
and the charge density $\rho(x^5)$ of such zero modes 
is not affected by the Wilson-line phase $e^{iax^5/R}$.
A phenomenological implication is that
the Dirac gaugino masses 
(at the GUT scale) are predicted to be universal, 
that is, common for gluino, wino and bino.

Note that
in the gGHU, the $SU(5)_G$ gauge symmetry is broken
by the Wilson-line VEV (\ref{WilsonLine}):
$e^{a\pi i}=+1$ for $SU(3)$ and $e^{a\pi i}=-1$ for $SU(2)$.
Therefore the above universality is not trivial at all,
and is specific 
to the case with large bulk-boundary mixing masses $\mu_{p}$.
It is interesting to extend the present work
to more general cases with a finite $\mu_p$
or the case without the boundary-localized fermion,
where nontrivial $a$-dependence is expected.

\red{%
Phenomenologically
it is very important 
whether the Dirac gaugino masses are universal or not,
both in the TeV scale scenario
and in the intermediate scale one that we mentioned in Introduction.
For instance,
to estimate the proton decay rate,
we have to know the gaugino mass spectrum
so as to determine the unification scale very accurately.
We hope to report on this point in the near future.
As for the proton decay,
its rate will depend on models of the flavor,
as was discussed in Ref.~\cite{gGHU-DTS};
it depends on where the first generation of quarks and leptons
reside in the extra dimensions.
A further study on these points will be desired.
}


An important assumption of the present work is that 
the bulk mass term of the messenger multiplets is 
the unique source of the $\Z2$ breaking (\ref{Eq:Z2breaking}).
In general,
once the $\Z2$ symmetry is broken by one sector,
one could introduce a similar $\Z2$-breaking effect
in the other sectors.
This includes a bare CS-like term,
and a $\Z2^{ex}\times\Z2^{5d}$-breaking mass term
for the Higgs hypermultiplets. 
Such term would contribute to a mass of the Higgs doublets,
destabilizing the EW scale.
A clever model building will be necessary
for the Higgs sector not to couple to 
the $\Z2$-breaking messenger sector:
such coupling, if exist, should be suppressed sufficiently.

Another issue related to
the discussion in Sect.~\ref{Sec:DTS&GCU} is how 
the $\Z2$ breaking affects
the correct pattern of the Wilson-line VEV (\ref{WilsonLine}):
a slight shift 
would spoil the DT splitting.
Actually this does not happen as can be seen from \Eq{Eq:KKfn}:
the KK mass spectrum of the bulk fermions is symmetric 
under the sign flip of the Wilson line, $a\rightarrow{}-a$.
Consequently
the Wilson-line VEV (\ref{WilsonLine})
is stable against the $\Z2$ breaking.
Note that this is true even for a finite $\mu_p$ case.


Finally we speculate about a possible origin 
of the proposed $\Z2$-breaking
by the bulk messenger mass term.
An idea is to apply the diagonal embedding method
to the $U(1)_D$ gauge group:
we introduce $U(1)_{D1} \times U(1)_{D2}$ gauge group
and identify its $\Z2^{ex}$-even combination 
as the $U(1)_D$ of the present model.
By denoting the odd combination by $U(1)_m$
and its gauge field by $B_M^{(-)}$,
we suppose that 
the extra-dimensional component $B_5^{(-)}$,
or its real scalar SUSY partner,
develops a nonvanishing VEV via some mechanism.
Then such a VEV will generate a fermion mass 
for $U(1)_m$-charged multiplets,
but not for $U(1)_m$-neutral fields.
This could explain our assumptions,
the presence of the bulk messenger mass term
and the absence of the bulk Higgs mass term.
This possibility might deserve further study.


\section*{Acknowledgment}


This work was supported in part 
by JSPS KAKENHI Grant No.~JP19K03865.
The work of O.S. was in part also supported 
by JSPS KAKENHI Grant No.~19K03860 and No.~21H00060.


\appendix

%
\section{Note on supersymmetric Lagrangian}
\label{Sec:SUSY}
%
%

\red{%
{}For completeness, 
we present the supersymmetric Lagrangian
of our gGHU setup for generating the mixed CS-like term,
recalling the 4D superfield formalism
for the 5D supersymmetric Lagrangian~\cite{HigherDimSUSY}.
}
%
As is well-known,
a 5D vector supermultiplet 
\red{(in the adjoint representation)} contains 
a 5D vector $A_M^a$ and a real scalar $\Sigma^a$ as the bosonic part.
The correspondence to 4D $\mathcal{N}=1$ superfields
is that the scalar component of a 4D chiral superfield 
$\Phi^a(x^M,\theta)$
is given by
$\phi^a(x^M)=\left(\Sigma^a+iA_5^a\right)/\sqrt{2}$.
%
\red{%
We identify their zero modes
with the scalar components, 
$\phi^a=\left(\sigma^a+i\pi^a\right)/\sqrt{2}$,
contained in Eq.~(\ref{Eq:adjoint}).
}

%
{}For the $SU(5)_1\times SU(5)_2$,
we introduce a $\Z2^{ex}$ pair of 5D vector supermultiplets,
$V^{(i)}(x^M,\theta,\bar{\theta})$ and 
$\Phi_{ad}^{(i)}(x^M,\theta)$ ($i=1,2$),
where 
%
\red{%
the $i$-th fields belong to the adjoint representation
of the $SU(5)_i$.
}
%
We have put the subscript ``ad'' 
to distinguish 
\red{the adjoint chiral multiplets from}
the hypermultiplets below.
The orbifold boundary conditions at $x^5=0$ are given by
\begin{align}
V^{(1)}(x^\mu,-x^5,\theta,\bar{\theta})
 ={}+V^{(2)}(x^\mu,x^5,\theta,\bar{\theta})
      \ , \qquad
\Phi_{ad}^{(1)}(x^\mu,-x^5,\theta)
 ={}-\Phi_{ad}^{(2)}(x^\mu,x^5,\theta)
      \ ,
\label{Eq:BC:vector}
\end{align}
while those at $x^5_1=\pi{}R$ 
will be affected
when we gauge away the Wilson line (\ref{WilsonLine}).

%
\subsection{Supersymmetric Lagrangian for messenger multiplets}
\label{Sec:SUSY:model}
%

Here we focus on the supersymmetric extension of the Lagrangians 
given in Eqs.~(\ref{Eq:Lbulk})--(\ref{Eq:Lboundary});
as for the gauge sector,
we will discuss the 5D supersymmetric CS term
in the next subsection.


In general, a 5D bulk hypermultiplet consists of a pair of 
4D $\mathcal{N}=1$ chiral superfields,
\begin{align}
\Phi_m(x^M,\theta)
&=   \phi_m(x^M)
     +\sqrt{2}\theta^\alpha\psi_{m\alpha}(x^M)
     +\cdots \ ,
\\
\Phi_m^c\fun{x^M,\theta}
 &=   \phi_m^c(x^M) 
     +\sqrt{2}\theta^\alpha\psi^c_{m\alpha}(x^M)
     +\cdots \ ,
 \end{align}
where higher components in superspace coordinates have been omitted.
In our case,
the messenger fermion fields $\Psi^{(i)}$ ($i=1,2$)
presented in Sect.~\ref{Sec:CSingGHU} are contained 
in a $\Z2^{ex}$ pair of bulk hypermultiplets,
$\Phi_m^{(i)}$ and $\Phi_m^{c(i)}$, 
with the identification
\begin{align}
\Psi^{(i)}
\ \equiv\ \left(\Psi_L^{(i)}, \Psi_R^{(i)}\right)^T
\ =\ \left(\psi_m^{(i)}, \bar{\psi}_m^{c(i)}\right)^T
  \ .
\label{Eq:Diracfermion}
\end{align}
Their representations under $SU(5)_1\times SU(5)_2\times U(1)_D$
are summarized in Table~\ref{tab:content:susy}.
At each boundary $x^5=x^5_{p=0,1}$,
we put the 4D chiral superfield $\phi_p$,
whose fermion component is the boundary-localized fermion $\psi_p$.
In the Table,
the representations under the unbroken gauge symmetry
are shown for these boundary-localized fields.

\begin{table}[tb]
\caption{
	Field contents of the supersymmetric gGHU setup 
        for generating the supersoft operators 
        in the Dirac gaugino models:
        the gauge vector multiplets and matter chiral multiplets
        as well as Higgs hypermultiplets are omitted.
}
\begin{center}
\begin{tabular}{c||cc|c|c}
\hline
	\makebox[12mm]{bulk fields} & 
	\makebox[30mm]{	$SU(5)_1{\times}SU(5)_2$ } & 
	\makebox[15mm]{	$SU(5)_{G}$ } & 
	\makebox[10mm]{ $U(1)_D$ } & 	
	\makebox[10mm]{	$P_5$ } 
	\\\hline 
	$\Phi_m^{(1)}(x^M,\theta)$ &
	$(\,\mathbf{R},\mathbf{1}\,)$ &	$\mathbf{R}$ & 
        $+q_D$ & $+$ 
	\\
	$\Phi_m^{(2)}(x^M,\theta)$ &
	$(\,\mathbf{1},\mathbf{R}\,)$ &	$\mathbf{R}$ & 
        $+q_D$ & $+$ 
	\\
	$\Phi_m^{c(1)}(x^M,\theta)$ &
	$(\,\mathbf{\bar{R}},\mathbf{1}\,)$ &
	$\mathbf{\bar{R}}$ & 
        $-q_D$ & $-$ 
	\\
	$\Phi_m^{c(2)}(x^M,\theta)$ &
	$(\,\mathbf{1},\mathbf{\bar{R}}\,)$ &
	$\mathbf{\bar{R}}$ & 
        $-q_D$ & $-$ 
        \\
	\hline \hline 
	boundary fields &
	localized position & 
	\makebox[15mm]{	$SU(5)_G$} & 
	\makebox[10mm]{ $U(1)_D$} & 	
	\makebox[10mm]{	$P_{ex}$ } 
        \\ \hline 
 $\phi_{p=0}(x^\mu,\theta)$ & 
        $x^5_0=\ 0\ $ &
	$\mathbf{R}$ & 
        $+q_D$ & $-$
 \\ 
 $\phi_{p=1}(x^\mu,\theta)$ & 
        $x^5_1=\pi{}R$ &
	$\mathbf{R}$ & 
        $+q_D$ & $-$
 \\\hline
\end{tabular}
\end{center}
\label{tab:content:susy}
\end{table}


The bulk hypermultiplets satisfy
the orbifold boundary conditions at $x^5_p$,
\begin{align}
\Phi_m^{(1)}(x^\mu,x^5_p{-}x^5,\theta)
 &={}+\Phi_m^{(2)}(x^\mu,x^5_p{+}x^5,\theta)
      \ ,
\nonumber\\
\Phi_m^{c(1)}(x^\mu,x^5_p{-}x^5,\theta)
 &={}-\Phi_m^{c(2)}(x^\mu,x^5_p{+}x^5,\theta)
      \ ,
\label{Eq:BC:hyper}
\end{align}
which lead to the zero modes 
in $\Phi_m^{(+)}$ and $\Phi_m^{c(-)}$
before we add the bulk mass term
and switch on the $SU(5)$-breaking 
by the Wilson line (\ref{WilsonLine}).

The supersymmetric action for the bulk messenger hypermultiplets
is given by
\begin{align}
S_{\mathrm{bulk}}
 =& \sum_{i=1,2}\int\!d^5x_{\,}d^4\theta
    \left( \Phi_m^{(i)\dagger} e^{-V} \Phi_m^{(i)}
          +\Phi_m^{c(i)} e^{+V} \Phi_m^{c(i)\dagger}
    \right)
\nonumber\\
 &+ \sum_{i=1,2}\int\!d^5x
    \left\{d^2\theta\,
            \Phi_m^{c(i)}\left(\widehat{D}_5+m_i\right)\Phi_m^{(i)}
           +c.c.
     \right\}
     \ ,
\label{Eq:Lbulk:SUSY}
\end{align}
where $\widehat{D}_5=D_5-\Sigma/\sqrt{2}$
is the 5th component of the gauge covariant derivative
that contains 
%
\red{
the real scalar $\Sigma$.
}
The bulk mass parameters $m_i$ are taken to be $m_1={}-m_2=m$
as we discussed in Sect.~\ref{Sec:CSingGHU}.

The supersymmetric action that contains 
the boundary Lagrangian (\ref{Eq:Lboundary}) is given by
%
%
\begin{align}
S_{\mathrm{boundary}}
 = &
     \int\!d^4x\sum_{p=0,1}
    \left[d^4\theta\,\phi_{p\,}^\dagger e^{-V}\phi_p
   +\left\{d^2\theta\,
     \sqrt{\mu_p}\,\phi_{p\,}\Phi_m^{c(-)}
           +c.c.
     \right\}
     \right]_{x^5=x^5_p}
     \ .
\end{align}
%
\red{%
In the limit of $\mu_p\rightarrow\infty$,
the bulk-boundary mixing terms force
the $\Phi_m^{c(i)}$ to obey Dirichlet boundary conditions,
leaving zero modes in the $\Phi_m^{(i)}$.
We will see this explicitly in Apppendix~\ref{Sec:limits}.
}
\subsection{Matching supersoft term to 5D supersymmetric CS term}
\label{Sec:Matching}
%


%
%

%
\red{Here we describe} 
how the supersoft term \red{(\ref{Eq:classic})}
written in terms of 4D $\mathcal{N}=1$ superspace language
\red{can be related}
to the bosonic CS-like term \red{(\ref{Eq:CS:goal})}
in the 5D Lagrangian.
%
\red{%
For this purpose,
we recall that
the supersymmetric CS term in the 5D Lagrangian
}
is given in terms of 4D $\mathcal{N}=1$ superspace 
notation of Ref.~\cite{HigherDimSUSY} by
\begin{align}
\mathcal{L}_{\mathrm{CS}}^{5d}
 &=  \int\!\mathrm{d}^2\theta
     \left\{
       \sqrt{2}\,\Phi W_\alpha W^\alpha
    -\frac{2}{3}\left(
        \partial_5 VD_\alpha V
       -VD_\alpha\partial_5V
     \right)W^\alpha
     \right\}
     +\Hc
\nonumber\\
 &\quad
    -\int\!\mathrm{d}^4\theta\,
     \frac{4}{3}\left[
        \partial_5V
       -\frac{1}{\sqrt{2}}\left(\Phi+\Phi^\dagger\right)
     \right]^3        
\nonumber\\
 &={}-\frac{1}{2}\,\epsilon^{LMNPQ}A_LF_{MN}F_{PQ}
     +\Sigma F^{MN}F_{MN}
     +2\Sigma\partial_M\Sigma\partial^M\Sigma
    +\cdots \ ,
\label{Eq:SUSYCS}
\end{align}
where $x^5$ dependencies are implicit, 
and
only the bosonic part is shown in the second expression.
Although the above expression is for a single gauge group,
it is straightforward to include 
mixed terms of several gauge groups, 
like those in the supersoft term (\ref{Eq:classic}).

Upon the reduction to 4D Lagrangian,
there appear the same zero mode wavefunctions
in the supersoft 
and the supersymmetric CS terms.
We see that the coefficient of 
the 4D supersoft operator (\ref{Eq:classic})
can be read off from the corresponding 5D term
of the form
\begin{align}
{}-\frac{1}{2}\,\epsilon^{LMNPQ}A_LF_{MN}F_{PQ}
\ ={}-A_5F_{\mu\nu}\widetilde{F}^{\mu\nu} \ .
\end{align}
Note that there appears a factor 2
when we extend the above to the mixed CS term,
as in \Eq{Eq:CS:goal}, 
but the same is true in both hand sides of \Eq{Eq:SUSYCS}.
We only have to take care of the normalization of 
the gauge couplings involved.


%
\section{Anomaly inflow and deformed Chern-Simons term}
\label{Sec:AnomalyInflow}
%

Here we recall some properties of 
anomalies on an $S^1/\Z2$ orbifold
and the relation to the CS term~\cite{%
AnomalyOnOrbifold1,AnomalyOnOrbifold2}. 
This includes some preliminaries 
for a discussion 
in Appendix~\ref{Sec:calculations}.
In this Appendix, we set the fermion charge to $1$ for simplicity.

Consider a theory on an $S^1/\Z2$ orbifold
with two fixed points $x^5_0=0$ and $x^5_1=\pi R$, 
and suppose that the theory possesses a chiral fermion zero mode
localized to one boundary at $x^5_0$.
Then the anomaly calculated from the fermion zero mode
is localized to $x^5_0$.
On the general ground, however, 
it can be shown~\cite{AnomalyOnOrbifold1,AnomalyOnOrbifold2}
that the anomalies should be localized equally 
on two boundaries $x^5_{p=0,1}$,
implying that 
there should be an anomaly inflow from $x^5_0$ to $x^5_1$.
Actually such an inflow is induced by the CS term:
It contains the gauge field $A_5$ explicitly,
and
its gauge variation $\delta_g$,
being a total derivative,
results in the surface terms.
These surface terms take the same form
as the localized anomaly,
and thus express the requisite inflow.

More generally one can consider a theory with a fermion zero mode
spread in the bulk with a charge density $\rho(x^5)$.
Then 4D current divergence suffers from anomaly\,\footnote{
This anomaly,
%
\red{%
induced by a single Weyl fermion,
}
is one-half of the one induced by a Dirac fermion. 
}
at each 4D slice in the bulk,
\begin{align}
\delta_g\Gamma_{\mathrm{ZM}}
\ =\ \int\!d^4x\!\int_{x^5_0}^{x^5_1}\!\!dx^5\,
     \rho(x^5)\,
     \frac{g^2}{16\pi^2}\,
     \alpha(x^5)\,F_{\mu\nu}\widetilde{F}^{\mu\nu}
     \ ,
\label{Eq:spreadanomaly}
\end{align}
where $\alpha$ is a transformation parameter of the fermion field.
In this case, the required anomaly inflow can be induced by a term
\begin{align}
\Gamma_{\mathrm{CS}}
\ =\  \int\!d^4x\!\int_{x^5_0}^{x^5_1}\!\!dx^5\,
      u(x^5)\,
      \frac{g^3}{16\pi^2}\,
      A_5 F_{\mu\nu}\widetilde{F}^{\mu\nu}
      \ ,
\label{Eq:deformedCS}
\end{align}
where $u(x^5)$ is an $x^5$-dependent coefficient function.
The gauge variation, 
$\delta_g A_5=(1/g)\partial_5\alpha(x^5)$,
gives, after partial integration,
\begin{align}
\delta_g \Gamma_{\mathrm{CS}}
\ ={}-\int\!d^4x\!\int_{x^5_0}^{x^5_1}\!\!dx^5\,
      \partial_5u(x^5)\,
      \frac{g^2}{16\pi^2}\,
      \alpha
      F_{\mu\nu}\widetilde{F}^{\mu\nu}
     +\int\!d^4x\!\left[
       u(x^5)\,
       \frac{g^2}{16\pi^2}\,
       \alpha
       F_{\mu\nu}\widetilde{F}^{\mu\nu}
      \right]_{x^5_0}^{x^5_1}
      \ .
\label{Eq:CS:variation}
\end{align}
We see that 
the spread anomaly (\ref{Eq:spreadanomaly}) in the bulk 
can be canceled if we require $\partial_5 u(x^5)=\rho(x^5)$,
supplemented with the boundary conditions
\begin{align}
u(x^5_1)
\ ={}-u(x^5_0)
\ =\ \frac{1}{2}\,\mathcal{A}
\ \equiv\ \frac{1}{2}
     \int_{x^5_0}^{x^5_1}\!\!dx^5\,\rho(x^5)
     \ ,
\label{Eq:BC:anomaly}
\end{align}
where $\mathcal{A}$($=1$) 
is the coefficient of a total anomaly in the 4D theory.

The CS term (\ref{Eq:deformedCS}) 
is a position-dependent term in the 5D Lagrangian,
called ``deformed'' Chern-Simons term~\cite{PiloRiotto2002}.
In Appendix~\ref{Sec:calculations}, 
we shall discuss a similar term in the gGHU setup.
The above discussion also makes it clear that
the coefficient of CS term on $S^1/\Z2$ orbifold
is not necessarily quantized, 
in contrast to the CS terms on a space without boundary.

The CS term on a consistent $S^1/\Z2$ orbifold
should be regarded as a $P_5$-even term 
in the 5D Lagrangian.
To see this point,
we go ``upstairs'', \textit{i.e.}, 
we work on the covering space $S^1$ 
with the coordinates ${}-\pi R<y<\pi R$.
Then we can regard the CS term (\ref{Eq:deformedCS}) 
as a $P_5$-even term
by extending the coefficient function $u(y)$
as an odd function of $y$.
For instance, a constant CS term on the $S^1/\Z2$ is actually
accompanied by 
$u(y)=\left(1/2\right)\varepsilon(y)$,
where $\varepsilon(y)=+1$ for $0<y<\pi R$
and $\varepsilon(y)=-1$ for $-\pi R<y<0$.
The situation is the same as for the kink mass term,
where the $P_5$-odd operator $\bar{\Psi}\Psi$ 
is accompanied by a $P_5$-odd function $m(y)=m\varepsilon(y)$.

We note also that
there are two kinds of the CS terms
in the gGHU setup in Sect~\ref{Sec:CSingGHU}.
The first type of the CS terms are
the usual one that are $P_{ex}$-even as well as $P_5$-even
in the above sense.
The other ones are
$P_{ex}$-odd and $P_5$-odd CS terms.
Adding a constant to the $u(y)$ over the $S^1$
corresponds to the latter terms accompanied by
the $P_5$-even extension of the $u(y)$.
It would be interesting 
if the quantization condition can be discussed for such terms.


%
\section{Calculations in a simplified setup}
\label{Sec:calc}
%


In this Appendix, 
we show some calculations in a simplified setup:
a five-dimensional 
$U(1)_1\times U(1)_2\times\Zex\times U(1)_D$ model.
%
\red{%
Table~\ref{tab:content:u1} shows the femion field contents 
relevent for the calcuation of the mixed CS-like term.
}
We introduce a $\Zex$ pair of bulk fermions 
$\Psi\supP1(q,0,q_D)$ and $\Psi\supP2(0,q,q_D)$,
where 
the $U(1)$ charges $\left(Q_1, Q_2, Q_D\right)$ are shown. 
The BCs around $x^5_p=p\pi{R}$ ($p=0,1$)
are $\Psi\supP1(x^5_p-x^5)={}-\gamma_5\Psi\supP2(x^5_p+x^5)$.
When each 5D Dirac fermion is decomposed into two Weyl spinors, 
%
\red{%
$\gamma_5\Psi\supP{i}_{R/L}={}\pm\Psi\supP{i}_{R/L}$,
the components that have zero modes are 
$\Psi\supP+_L$ and $\Psi\supP-_R$.
}
In addition, a 
left-handed fermion $\psi_{L}^{p=0,1}(q,-,q_D)$ is put at each boundary.
Note that for the boundary-localized fermions,
only the quantum numbers of the unbroken symmetry 
$(U(1)_V, \Zex, U(1)_D)$ are given,
since the $U(1)_A$ corresponding to $Q_1-Q_2$ is broken there.

\begin{table}[tb]
\caption{
	Field contents of the simplified model:
}
\begin{center}
\begin{tabular}{c||cc|c|c}
\hline
	\makebox[12mm]{bulk fields} & 
	\makebox[30mm]{	$U(1)_1{\times}U(1)_2$ } & 
	\makebox[30mm]{	$U(1)_V{\times}U(1)_A$ } & 
	\makebox[10mm]{ $U(1)_D$ } & 	
	\makebox[10mm]{	$P_5$ } 
	\\\hline 
	$\Psi_L^{(1)}(x^M)$ &
	$(\,q,0\,)$ & $(\,q,+q\,)$ & $q_D$ & $+$ 
	\\
	$\Psi_L^{(2)}(x^M)$ &
	$(\,0,q\,)$ & $(\,q,-q\,)$ & $q_D$ & $+$ 
	\\
	$\Psi_R^{(1)}(x^M)$ &
	$(\,q,0\,)$ & $(\,q,+q\,)$ & $q_D$ & $-$ 
	\\
	$\Psi_R^{(2)}(x^M)$ &
	$(\,0,q\,)$ & $(\,q,-q\,)$ & $q_D$ & $-$ 
        \\
	\hline \hline 
	boundary fields &
	localized position & 
	\makebox[30mm]{	$U(1)_V$ } & 
	\makebox[10mm]{ $U(1)_D$ } & 	
	\makebox[10mm]{	$P_{ex}$ } 
        \\ \hline 
 $\psi_L^{p=0}(x^\mu)$ & 
        $x^5_0=\ 0\ $  & $q$ & $q_D$ & $-$
 \\ 
 $\psi_L^{p=1}(x^\mu)$ & 
        $x^5_1=\pi{}R$ & $q$ & $q_D$ & $-$
 \\\hline
\end{tabular}
\end{center}
\label{tab:content:u1}
\end{table}


%
\subsection{KK spectrum}
\label{Sec:KK}
%

The relevant bulk Lagrangian is given by Eq.~(\ref{Eq:Lbulk}), 
with a modification
\begin{align}
D_M
 &=  \del_M
     -ig\left(
        Q_1A\supP1_M+Q_2A\supP2_M
       \right)
     -i{\g5D} Q_D B_M
\nn\\
  &=  \del_M
      -i\,\frac{g}{\sqrt{2}}\left(Q_1+Q_2\right)A^{(+)}_M
      -i\,\frac{g}{\sqrt{2}}\left(Q_1-Q_2\right)A^{(-)}_M
      -i{\g5D}Q_D B_M
       \ , 
\end{align}
where
$\g5D$ is the gauge coupling constant
of the $U(1)_D$ gauge field $B_M$.
Note that
the gauge couplings of 
$A_M^{(\pm)}=(A_M^{(1)}\pm A_M^{(2)})/\sqrt{2}$
are normalized to be $g/\sqrt{2}$.
We introduce the general bulk mass terms allowed by 
the gauge symmetry and the $\Z2^{comb}$-twist $P^5P_{ex}$;
\bequ
{}-\mathcal{L}_{\mathrm{mass}}^{\mathrm{bulk}}
 = m_+\left\{\bar\Psi\supP1\Psi\supP1+\bar\Psi\supP2\Psi\supP2\right\}
  +m_-\left\{\bar\Psi\supP1\Psi\supP1-\bar\Psi\supP2\Psi\supP2\right\}
   \ .
\label{Eq:bulkmass:general}
\eequ
The $P_{ex}$-even mass should have a kink profile, 
which we take to be a constant value $m_+$ 
in the fundamental region $0<x^5<\pi{}R$.
On the other hand,
the $P_{ex}$-odd mass $m_-$ is a constant over $S^1$.
%
%
Working in the fundamental region of $S^1/\Z2$,
we have $m_1=m_++m_-$ and $m_2=m_+-m_-$.

The bulk EOMs for the Weyl spinor 
fields $\Psi_{\chi=L,R}\supP{i}(x^\mu,x^5)$ are
\beqn
 &&i\sla D\,\Psi\supP{i}_R=\left(-D_5+m_i\right)\Psi\supP{i}_L \ , \\
 &&i\sla D\,\Psi\supP{i}_L=\left(+D_5+m_i\right)\Psi\supP{i}_R \ . 
\eeqn 
We decompose these fields 
by using the mode functions with a 4D mass $\mKK$, 
$\xi_{\chi=L,R}\supP{i}(x^5)$, satisfying
\beqn
 &&\mKK \xi\supP{i}_R\fun{x^5}
   = \left(-D_5+m_i\right)\xi\supP{i}_L\fun{x^5}\ , 
\label{Eq:bulkEOMR}\\
 &&\mKK \xi\supP{i}_L\fun{x^5}
   = \left(+D_5+m_i\right)\xi\supP{i}_R\fun{x^5} \ , 
\label{Eq:bulkEOML}
\eeqn
from which we have
\bequ
 \mKK^2\xi\supP{i}_\chi(x^5)
 = \left(-D_5^2+m_i^2\right)\xi\supP{i}_\chi(x^5)
   \ . \qquad \left( \ \chi=L,\ R\ \right)
\label{Eq:bulkEOMLR}
\eequ
%
The general solutions 
are given by 
\bequ
 \xi\supP{i}_\chi(x^5)
 = e^{ia_ix^5/R}
   \Ls c_{\chi1}\supP{i}\,e^{\sqrt{m_i^2-\mKK^2}x^5}
      +c_{\chi2}\supP{i}\,e^{-\sqrt{m_i^2-\mKK^2}x^5}
   \Rs
   \ . 
\label{Eq:profile}
\eequ
Here we have introduced the Wilson-line phase,
$a_1={}-a_2=a$,
where 
\bequ
a
\ \equiv\ R\frac{qg}{\sqrt2}\VEV{A\supP-_5}
          \ .
\label{Eq:background}
\eequ
The coefficients 
$c_{L\ell}\supP{i}$ and $c_{R\ell}\supP{i}$ ($\ell=1,\,2$) 
are related with each other 
via the EOMs (\ref{Eq:bulkEOMR}) and (\ref{Eq:bulkEOML}). 

The KK spectrum is determined from the BCs, 
which are affected by the boundary fields. 
At each boundary $x^5_{p=0,1}$,
we introduce a bulk-boundary mixing mass term 
(\ref{Eq:Lboundary}), 
where $X\supP{\pm}\equiv(X\supP1 \pm X\supP2)/\sqrt2$.
Then the EOM of $\bar\Psi_R\supP-$ is modified into
\bequ
 i\sla D\,\Psi\supP{-}_R
 ={}-\del_5\Psi\supP{-}_L
 +\sqrt{\mu_p}\,\psi_L^p\,2\delta(x^5-x^5_p) + \cdots
     \ , 
\eequ
where we omit terms finite at $x^5=x^5_p$.
Integrating it over a tiny region 
$[x^5_p-\epsilon,x^5_p+\epsilon]$, we get 
\bequ
 (-1)^p \Psi\supP{-}_L(x_p^{5\epsilon})=\sqrt{\mu_p}\psi^p_L  
 \ . \qquad
 \left(\ x_p^{5\epsilon}\equiv x^5_p+(-1)^p\epsilon\ \right)
\eequ
Now, we find from the KK decomposition, 
with the help of the EOM of $\bar\psi_L^p$, 
the $P_5$-oddness of 
$\Psi\supP{-}_L(x^5_p+\epsilon)=-\Psi\supP{-}_L(x^5_p-\epsilon)$, 
and the continuity of $\Psi_R\supP-$, that 
\bequ
    (-1)^p \xi\supP{-}_L(x_p^{5\epsilon}) 
\ =\ \frac{\mu_p}{\mKK}\,\xi_R\supP-(x^5_p)
\ \sim\ 
     \frac{\mu_p}{\mKK}\,\xi_R\supP-(x_p^{5\epsilon})
   \ .
\label{Eq:BCs-}
\eequ
Later we shall treat the massless modes separately
since this condition is ill-defined at $\mKK=0$.
Working in the fundamental region $0\le x^5 \le\pi R$,
we take the limit $\epsilon\to0$ hereafter.

We are particularly interested in the limit 
$\mu_p\gg1/R$ (and also $\mu_p\gg m_i$).
In this large $\mu_p$ limit, 
we see from \Eq{Eq:BCs-} that
the lower-lying KK modes of $\Psi_R\supP-$ (those with $\mKK\ll \mu_p$)
effectively obey the Dirichlet BC,
while the BC of $\Psi_L\supP-$ is determined by the EOMs.
As for $\Psi\supP+$, there are no bulk-boundary mixing mass, 
and their BCs correspond to taking $\mu_p=0$. 
Recalling that $\Psi\supP+$ obeys the BCs 
opposite to those of $\Psi\supP-$, we see that 
\bequ
 \Psi\supP{+}_R(x_p)=0 \ . 
\label{Eq:BCs+}
\eequ

Inserting the general solutions (\ref{Eq:profile})
into the four BCs (\ref{Eq:BCs-}) and (\ref{Eq:BCs+}), 
we have four equations which are linear and homogeneous 
in the four parameters $c_{L\ell}\supP{i}$. 
Then, nontrivial solutions are obtained 
for specific values of the KK mass $\mKK$ that 
make the determinant of the $4\times4$ coefficient matrix 
of the simultaneous equations vanishing. 
A straightforward calculation shows that
the KK mass spectrum $\mKK=\mKK_n$
is given by the zeros of the function,
\begin{align}
N(\mKK;a)
\ =\ \frac{8\omega_1\omega_2}{\mKK^2}
     \Bigl[\cos(2\pi a)-N_c(\mKK)\Bigr]  
     \ ,
\label{Eq:KKfn}
\end{align}
where
 we have defined
$\omega_i(\mKK)=\sqrt{\mKK^2-m_i^2}$ ($i=1,2$) and 
\begin{align}
 N_c(\mKK)
 =&\cos\Ls\omega_1\pi R\Rs \cos\Ls\omega_2\pi R\Rs 
\nn\\
  &+\frac{\mu_0-\mu_1}{ \omega_1}
    \sin\Ls\omega_1\pi R\Rs
    \cos\Ls\omega_2\pi R\Rs
  +\frac{\mu_0-\mu_1}{ \omega_2}
    \cos\Ls\omega_1\pi R\Rs 
    \sin\Ls\omega_2\pi R\Rs
\nn\\
  &-\frac{(\omega_1^2+\omega_2^2)
          +\left[2\mu_0-(m_1+m_2)\right]\left[2\mu_1-(m_1+m_2)\right]
          }{2\omega_1\omega_2}
          \sin\Ls\omega_1\pi R\Rs 
          \sin\Ls\omega_2\pi R\Rs
  \ .
\end{align}
The spectrum in the KK tower can be read off
from the last parenthesis in \Eq{Eq:KKfn},
while the overall factor should be treated carefully, 
especially for $\mKK=0$, since the BCs may be ill-defined. 

%
\subsection{Limit of large bulk-boundary mixing}
\label{Sec:limits}
%

Here we consider some limiting cases;
we turn off the Wilson line, $a=0$, for simplicity.

First, let us consider the limit $\mu_p=0$
so that boundary fields decouple from the bulk ones.
In this case,
we can explicitly check that
in the presence of the $P_{ex}$-odd mass term,
$m_-=(m_1-m_2)/2\neq0$,
the bulk fermions have no massless mode.
Indeed, 
the EOMs (\ref{Eq:bulkEOMR}) and (\ref{Eq:bulkEOML}) tell us 
that the zero modes should have a profile 
\bequ
  \xi\supP{i}_{L 0}(x^5) = c_L\supP{i}e^{+m_ix^5} \ , \qquad
  \xi\supP{i}_{R0}(x^5) = c_R\supP{i}e^{-m_ix^5} \ .
\eequ
At $x^5=x^5_p$ ($p=0,1$),
the $P_5 P_{ex}$-odd fields
$\Psi\supP+_R$ and $\Psi\supP-_L$ obey the Dirichlet BCs
\bequ
 c\supP1_R e^{-m_1x^5_p} + c\supP2_R e^{-m_2x^5_p}=0 \ ,\qquad
 c\supP1_L e^{m_1x^5_p} - c\supP2_L e^{m_2x^5_p}=0 \ .
\eequ
which forces $c\supP{i}_\chi=0$ (if $m_1\neq m_2$).

We are mainly interested in the opposite limit 
in which the bulk-boundary mixing masses $\mu_{p=0,1}$ 
are much larger than 
the compactification scale $1/R$ 
and the bulk mass parameter $m_i$.
In this limit, 
the BCs of $\Psi\supP-$ changes effectively 
so that 
both of
$\Psi\supP+_R$ and $\Psi\supP-_R$ obey the Dirichlet BCs:
\bequ
 c\supP1_R e^{-m_1x^5_p} + c\supP2_R e^{-m_2x^5_p}=0 \ ,\qquad
 c\supP1_R e^{-m_1x^5_p} - c\supP2_R e^{-m_2x^5_p}=0 \ ,
\eequ
which forces $c\supP{i}_R=0$, 
but $c\supP{i}_L$ remains unconstrained. 
This means that there appear two zero modes 
$\xi\supP{1}_{L0}\propto e^{m_1x^5}$ and 
$\xi\supP{2}_{L0}\propto e^{m_2x^5}$, 
which behave differently for $m_1\neq m_2$.
In particular, when $m_1m_2<0$, namely $\abs{m_+}<\abs{m_-}$, 
the zero mode
$\xi\supP{2}_{L0}$ is localized towards the boundary 
opposite to the one around which $\xi\supP{1}_{L0}$ is localized. 

In the above limit $\mu_p\gg1/R$ and $\mu_p\gg \abs{m_i}$, 
Eq.~(\ref{Eq:KKfn}) reduces to
\bequ
N(\mKK;a)
\ \sim\
      \frac{16\mu_0\mu_1}{\mKK^2}\,
      \sin\Ls\omega_1\pi R\Rs
      \sin\Ls\omega_2\pi R\Rs
      \ , 
\eequ
from which we see that
the nonzero KK masses are given by
$\mKK_n^2=m_i^2+(n/R)^2$ ($n=1,2,\cdots$).
We note that
this is true even in the presence of the Wilson line:
The KK spectrum becomes independent of the Wilson line 
in this large $\mu_p$ limit.

If we consider the further limit $\abs{m_i}\gg1/R$,
all the nonzero KK mass $\mKK_n$ become much larger 
than the compactification scale.
Then we can integrate out the KK modes within the 5D picture,
which results in the effective CS-like 
term~\cite{AnomalyOnOrbifold1,AnomalyOnOrbifold2}. 
Away from such limit,
it may not be appropriate to integrate out KK modes 
within the 5D theory;
the CS-like term may be ill-defined 
from the 5D point of view. 
In the 4D effective theory, however,
we can still integrate out KK modes,
which results in the effective term (\ref{Eq:classic})
for the corresponding zero modes.

%
\subsection{Calculating the coefficient of CS-like term}
\label{Sec:calculations}
%

Here we calculate the coefficient of 
the mixed CS-like term (\ref{Eq:CS:goal})
in $m_+=0$ case.
For notational simplicity we denote $m_-$ just as $m$ hereafter.
In the simplified setup,
we can directly calculate it
without using \Eq{Eq:CSingGHU}.
We define the $U(1)^{(-)}$ current $J_M^{(-)}$
normalized according to
\begin{align}
S_{\mathrm{int}}
\ =\ \int\!d^4x\!\int_0^{\pi R}\!\!dx^5\,
     \frac{g}{\sqrt{2}}\,A_M^{(-)}J^M_{(-)}+\cdots
     \ .
\end{align}
%
Then, with the normalized zero mode wavefunctions 
\bequ
\xi_{L0}\supP{i}\fun{x^5}
\ =\ \sqrt{\frac{\modd}{\sinh(\modd\pi R)}}\,
     e^{\pm\modd\left(x^5-\frac{\pi R}{2}\right)}
     \ ,
\label{Eq:ZM:normalized}
\eequ
the four-divergence of the $U(1)\supP-$ current
spreads in the bulk according to\,\footnote{%
An extra minus sign comes from $\gamma_5=-1$ 
for left-handed zero modes.
}
\beqn
 \rho(x^5)
  ={}-
       q\abs{\xi_{L0}\supP{1}(x^5)}^2
      +q\abs{\xi_{L0}\supP{2}(x^5)}^2
\ =\ \frac{2q\,\modd}{\sinh(\modd\pi R)}
     \sinh\!\left[2\modd\Ls\frac{\pi R}2-x^5\Rs\right]
     \,. 
\label{Eq:rho0(x^5)}
\eeqn
The mixed CS-like term is to be generated to cancel this spread anomaly.

Assuming a background gauge field 
homogeneous with respect to the fifth direction,
let us write the requisite CS-like term in the form
\beqn
\Gamma_{\mathrm{CS}}
\ =\ 
     \frac{2(qg/\sqrt{2})^2q_D\g5D}{16\pi^2}
     \int\!d^4x\!\int_0^{\pi R}\!\!dx^5\,
     u\fun{x^5}A_5^{(-)}F_{\mu\nu}^{(+)}\widetilde{F}_D^{\mu\nu}(x^5)
     \ ,
\label{Eq:CSlike}
\eeqn
where 
$u(x^5)$ 
represents a possible $x^5$ dependence.
Note that we have put a factor $2$
for the mixed CS-like term.


The coefficient function $u\fun{x^5}$ can be determined
in several ways.
Our method here is inspired by the fact that
the distribution $\rho(x^5)$ is odd under the reflection 
around the midpoint $x^5=\pi R/2$:
We first pick up $0<y<\pi R/2$ 
and consider an inflow from a point $x^5=y$ on the one side 
to the point $x^5=\pi R-y$ on the opposite side. 
Such an inflow is induced by the CS-like term 
restricted to this interval 
$\left[y,\,\pi R-y\right]\equiv{}I_y$,
\begin{align}
L_{\mathrm{CS}}\funal{I_y}
 \equiv \int_0^{\pi R}\!dx^5\,\Theta_{y}(x^5)\,
        A_5^{(-)}F_{\mu\nu}^{(+)}\widetilde{F}_D^{\mu\nu}(x^5)
 \ .
\label{eq:inflow:restricted}
\end{align}
Here
$\Theta_{y}(x^5)$ is a rectangular support function 
of the interval $I_y$
and is given by
\bequ
 \Theta_{y}(x^5)
 = \Lm
 \begin{array}{l}
    1\qquad {\rm for}\quad x^5\in\left[y,\,\pi R-y\right]
 \\ 0\qquad{\rm otherwise}
 \end{array}\RR \ .
\label{Eq:supportFn}
\eequ
%
Then the gauge variation $\delta_g$ with a parameter $\alpha(x^5)$
gives, after the partial integration, 
an inflow from the point $y$ to $\pi R -y$ as
\begin{align}
 \delta_g L_{\mathrm{CS}}\funal{I_y}
 = \int_0^{\pi R}\!dx^5\,\Theta_{y}(x^5)
    \left(\frac{\sqrt{2}}{g}\del_5\alpha\right)
     F_{\mu\nu}^{(+)}\widetilde{F}_D^{\mu\nu}(x^5)
  = \frac{\sqrt{2}}{g}\left[
     \alpha F_{\mu\nu}^{(+)}\widetilde{F}_D^{\mu\nu}
    \right]_{x^5=y}^{\pi R-y}
      \ .
\end{align}
Since the anomaly spreads as $\rho(x^5)$, 
it is canceled by integrating \Eq{eq:inflow:restricted}
with the weight $\rho(x^5)$ as
\begin{align}
   \int_0^\frac{\pi R}2dy\,\rho(y)
   L_{\mathrm{CS}}\funal{I_y}
 &= 
    \int_0^\frac{\pi R}2dy\,\rho(y)
    \int_0^{\pi R}\!dx^5\,\Theta_{y}(x^5)
         A_5^{(-)}F_{\mu\nu}^{(+)}\widetilde{F}_D^{\mu\nu}(x^5)
\nonumber\\
 &= \int_0^{\pi R}\!dx^5\,
         q\,
         u\fun{x^5}A_5^{(-)}F_{\mu\nu}^{(+)}\widetilde{F}_D^{\mu\nu}(x^5)
   \ .
\label{Eq:CSlike:tmp}
\end{align}
Now, the integration can be carried out 
for the function $\rho(x^5)$ in Eq.~(\ref{Eq:rho0(x^5)})
with the result
\begin{align}
u\fun{x^5}
 =
     \int_0^\frac{\pi R}2dy\,
     \frac{1}{q}
     \rho(y)\Theta_{y}(x^5)
 =
    \,\frac{2\,
            \sinh\Ls \modd x^5\Rs
            \sinh\left[\modd \Ls{\pi R}-x^5\Rs\right]%
          }{\sinh(\modd\pi R)}
    \ . 
\label{Eq:u}
\end{align}
In this way we obtain the mixed CS-like term,
Eq.~(\ref{Eq:CSlike}) with the coefficient function (\ref{Eq:u}).


In view of the preliminary discussion 
given in Appendix~\ref{Sec:AnomalyInflow},
an alternative way to determine $u(x^5)$
would be to solve $\partial_5 u(x^5) = \rho(x^5)/q$,
with a suitable boundary condition.
It appears that
the absence of the surface terms 
requires $u(x^5_p) = 0$ at each boundary,
since we are considering a theory with no net anomaly.
This is not the case, however.
Since we are considering the gauge variation of 
the $U(1)^{(-)}$, which is broken by the orbifold BCs,
the parameter $\alpha(x^5)$ vanishes at the boundaries.
Consequently no boundary condition is required 
and thus we are left with an undetermined integration constant
for $u(x^5)$.
Note that such an integration constant would correspond
to adding a bare mixed CS-like term.

Finally we compute 
the corresponding term in the 4D effective theory.
To this end,
we keep only the zero mode part of each field. 
The zero mode wavefunctions, $1/\sqrt{\pi R}$,
relate the 5D gauge coupling constants to the 4D ones,
$g_G=g/\sqrt{2\pi R}$ for the unified gauge coupling constant 
and $\gD=\g5D/\sqrt{\pi R}$ for $U(1)_D$.
Thus integrating the 5D term over $0\leq x^5\leq\pi R$ 
yields the 4D term
\bequ
\mathcal{L}_{\mathrm{CS}}^{4d}
\ =\ 
     \frac{2q^2g_G^2q_Dg_{D}}{16\pi^2}\,
     \pi R\,
     f(\modd\pi R)\,
     A_5^{(-)}F_{\mu\nu}^{(+)}\widetilde{F}_D^{\mu\nu}(x)
     \ , 
\label{Eq:4D-CSterm:U1}
\eequ
where 
all the fields are 4D ones, and
the coefficient function $f(z)$ is defined by
\bequ
 f(z)
 \equiv\frac{1}{\tanh{z}}-\frac{1}{z}
 = \Lm\begin{array}{rcc}
    z/3 &{\rm for}& z\ll 1 \\
    1   &{\rm for}& z\gg 1
 \end{array}\RR
 \ .
\label{Eq:coefficientfunction}
\eequ
We recall that the above result 
is obtained from the messenger fermion multiplet
$\Psi^{(1)}(q,0,q_D)$ and 
$\Psi^{(2)}(0,q,q_D)$ of the $U(1)_1\times U(1)_2\times U(1)_D$.
After generalizing to $SU(5)_1\times SU(5)_2\times U(1)_D$ case,
we obtain the announced result (\ref{Eq:4D-CSterm:SU5}).

\end{document}